\documentclass[%
 reprint,
 amsmath,amssymb,
 aps,
prb,
]{revtex4-2}

\usepackage{graphicx}
\usepackage{dcolumn}
\usepackage{bm}
\usepackage{amsmath}
\mathchardef\mhyphen="2D 
\usepackage{float}
\usepackage[dvipsnames]{xcolor}
\usepackage[switch]{lineno}

\begin{document}

\preprint{APS/123-QED}

\title{The Mixing Thermodynamics and Local Structure of High-entropy Alloys from Randomly Sampled Ordered Configurations}
\author{Andrew Novick}
\affiliation{Colorado School of Mines}

\author{Quan Nguyen}
\affiliation{Washington University in St. Louis}

\author{Roman Garnett}
\affiliation{Washington University in St. Louis}

\author{Eric Toberer}
\affiliation{Colorado School of Mines}

\author{Vladan Stevanovi\'c}
\email{vstevano@mines.edu}
\affiliation{Colorado School of Mines}


\date{\today}

\begin{abstract}
A general method is presented for modeling high entropy alloys as ensembles of randomly sampled, ordered configurations on a given lattice. Statistical mechanics is applied \textit{post hoc} to derive the ensemble properties as a function of composition and temperature, including the free energy of mixing and local structure. Random sampling is employed to address the high computational costs needed to model alloys with a large number of components. Doing so also provides rigorous convergence criteria, including the quantification of noise due to random sampling, and an estimation of the number of additional samples required to lower this noise to the needed/desired levels. This method is well-suited for a variety of cases: i) high entropy alloys, where standard lattice models are costly; ii) “medium” entropy alloys, where both the entropy and enthalpy play significant roles; and iii) alloys with residual short-range order. Binary to 5-component alloys of the group-IV chalcogenides are used as case examples, for which the predicted miscibility shows excellent agreement with experiment.
\end{abstract}

\maketitle

\section{\label{sec:level1}Introduction}
From a materials design perspective, alloying provides a continuous composition space over which multiple conflicting properties can be simultaneously optimized. This flexibility has allowed for semiconducting alloys to have a wide variety of applications, including use in LEDs, solar cells, batteries, and thermoelectric generators \cite{schnepf2020utilizing,king2007solaralloy,sixjunction2020,Witting2019HEalloy}. Increasingly, this desire for multi-property optimization encourages the exploration of high-entropy alloys \cite{oses2020high}.
The complexity of these chemical spaces necessitates computational guidance with respect to  alloy stability, local and long-range structure, and material properties. For alloys with fewer components, \textit{ab-initio} methods such as Special Quasi-random Structures, Cluster Expansion, and the Independent Cell Approximation are effective in guiding experimental efforts \cite{zunger1990special,ferreira1989first,independent-supercell-alloy-2016,pomrehn2011entropic}. 
Each of these approaches faces significant challenges when moving to higher dimensions; this work presents how the Independent Cell Approximation can be modified to efficiently explore high-dimensional alloys.  

The Special Quasi-random Structure (SQS) approach uses a single supercell whose atomic arrangement is optimized to emulate a fully disordered (random) alloy structure within the constraints of periodic boundary conditions \cite{zunger1990special}.
Alloying on both sub-lattices of an ionic system can result in short-range order due to the diversity of constituent atom sizes and interactions; as such the complete disorder approximation of SQS overestimates the enthalpy of mixing \cite{schnepf2020utilizing,lany-anti-SQS}. Further, no information on entropy is provided--the analytic approach associated with SQS overestimates entropy by assuming a fully random alloy \cite{oses2020high}, disregarding any possibility for correlated disorder or short-range order \cite{keen2015crystallography,simonov2020designing}. 

Despite these stability challenges, the use of a single supercell allows for the full suite of plane-wave based DFT methods to predict material properties. However, the extremely  large supercells required for complex alloys quickly become computationally costly since DFT scales with the number of atoms as $O(N^3)$. 


Unlike SQS, Cluster Expansion (CE) does not assume complete disorder; instead, a model Hamiltonian is fit with a series of total-energy calculations on a number of small supercells of varying configurations and compositions. An accurate model Hamiltonian is a powerful tool--with it, the calculation of arbitrarily large supercells of varying configurations becomes almost effortless. Coupled with Monte Carlo, a model Hamiltonian can be used to predict the order-disorder behavior across the modeled alloy composition space, all as a function of synthesis-temperature \cite{cordell2021probing,wolverton1998first,van2002automating}. This is particularly useful for studying systems with short-range order. Still, the number of interaction parameters required to fit a cluster expansion grows with the number of alloyed elements, necessitating more calculations. The large number of calculations makes the method costly--and even prohibitive--for high entropy spaces. Furthermore, the model Hamiltonian does not provide predictions on the overall structure of the alloy. In the field of semiconducting alloys, the extended strain fields and distortions arising from alloying are important for the ways in which they affect transport properties and band dispersions \cite{zunger1983structural, ortiz2015effect, gurunathan2020alloy, gurunathan2022mapping}.

\begin{figure}
\includegraphics[width =0.9 \linewidth]{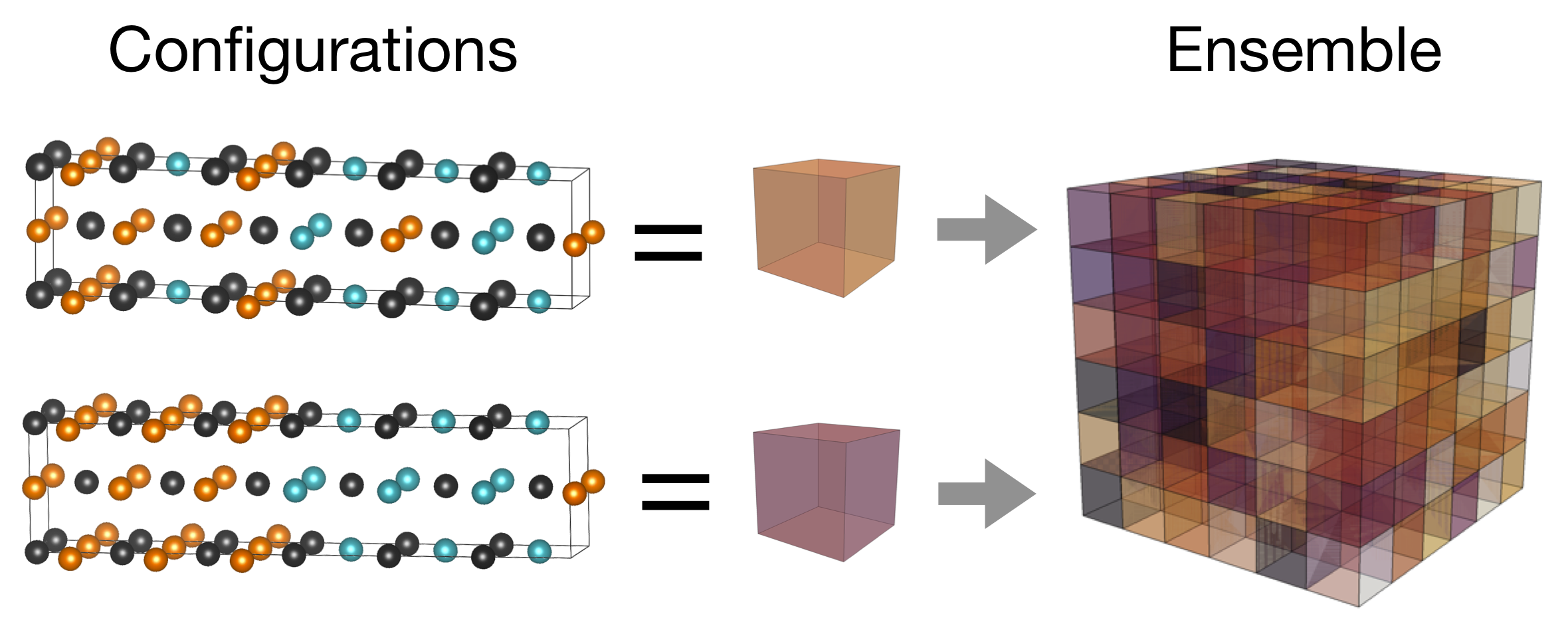}
\caption{\label{fig:CUBE} Alloys are represented as an ensemble of  configurational states within the Independent Cell Approximation.
\textit{Post hoc} application of statistical mechanics then predicts the alloy's local structure, thermodynamic stability and material properties.  }
\end{figure}

Within the independent cell approximation (ICA), the material is modeled as an ensemble of ordered configurational states, as illustrated in Fig.~\ref{fig:CUBE} \cite{pomrehn2011entropic,independent-supercell-alloy-2016,aflowpocc}. Specifically, calculations are first run on multiple supercells of varying configurations at or near a single composition.
The probability of each configurational state is then calculated using statistical mechanics. 
Finally, the material properties are calculated by taking the weighted average over the configurations. 
In contrast to CE, no model Hamiltonian is fit; the ensemble properties are derived directly from the cells within the ensemble. 
By avoiding the rigid lattice assumed by CE, the resulting ensemble averages can describe local structural distortions arising from disorder (e.g., deviations in bond lengths, bond angles, and local volumes). 
The ICA has been successfully used in a variety of applications: alloying \cite{aflowpocc,aflowlvtc,aflowefa,independent-supercell-alloy-2016,woods2022role},  composition disorder \cite{pomrehn2011entropic}, spin disorder \cite{gorai2016thermoelectricity}, metastable polymorphs \cite{stevanovic2016sampling,jones2017polymorphism,therrien2021metastable}, and glasses \cite{jones2020glassy,perim2016spectral}.

In the context of alloys, there has been considerable diversity in how  ICA is implemented. 
Key differences have involved supercell size, sampling of the configurational space, and the treatment of configurational entropy.   
An early example of ICA by Jiang et al.~sought to replicate large SQS of inverse spinels with relatively few smaller supercells; the contributions of these supercells were weighed such that the correlation function of the ensemble resembled a fully random alloy \cite{independent-supercell-alloy-2016}. Using just two 28-atom supercells was sufficient to determine an ensemble energy that agreed with both a 168-atom SQS and a cluster expansion (with a cross validation error of 3.4 meV/formula unit) . By focusing on matching a large SQS cell, this embodiment of ICA is not able to interrogate the temperature dependence of local structure and properties. \par

In addition to reproducing SQS results, ICA also has the capability of measuring configurational entropy. Leder et al.~fit various ICA configurations to a cluster expansion in order to estimate the solubility as a function of temperature \cite{aflowlvtc}; however,  cluster expansion in high entropy spaces is computationally expensive.
Sarker et al.~developed an  approximation for the  configurational entropy that directly uses ICA results termed the ``entropy forming ability'' (EFA). This metric is based on the spread of the supercell energies. 
In this work, the metric is applied to 56 high entropy carbides (V$_{0.2}$W$_{0.2}$X$_{0.2}$Y$_{0.2}$Z$_{0.2}$C); experimental formation of homogeneous alloys was found to correlate with EFA  magnitude, regardless of the enthalpy of mixing. 
As these carbides all have the same configurational entropy in the high-temperature limit, this work highlights the importance of considering the configurational entropy at finite temperatures.  

This paper addresses opportunities for improving the ICA, with a focus on ($i$) quantitatively determining  configurational entropy, and thus, the free energy convex hull; ($ii$) accurately modeling local structural distortions; and ($iii$) exploring high-dimensional chemical spaces. 
In all cases, this is made computationally tractable  by randomly sampling a portion of the configurational space, thus allowing for the use of larger, more disordered supercells within a first-principles framework. 

First, PbSe$_{0.5}$Te$_{0.5}$ and PbS$_{0.5}$Te$_{0.5}$ are used as a way of illustrating the methodology. These compositions are chosen because they have sufficiently few configurations. Therefore, ensemble statistics can be derived from completely sampling all possible configurations for a given supercell size. Convergence tests are then run to determine the necessary supercell size and number of configurations needed to capture the configurational thermodynamics and local structure of these IV-VI alloys.  
With complete sampling as a benchmark, random sampling is then conducted on the same systems with far less computational cost and little added uncertainty. The uncertainty due to random sampling is then derived by using the Central Limit Theorem \cite{CLT,CLT-History} in order to quantify the robustness of the result. With a rational approach to random sampling in hand, this case example is extended to the pseudo-ternary, PbS$_{x}$Se$_{y}$Te$_{1-x-y}$.  The temperature-dependent phase diagram is calculated and matches well with experiment. Finally, the computational tractability of this method in high-entropy spaces is illustrated through the exploration of the quintary, (Pb$_{1-x}$Ge$_{x}$)(Te$_{y}$Se$_{z}$S$_{1-y-z}$).  

\begin{figure*}[t!]
\includegraphics[width=0.95\linewidth]{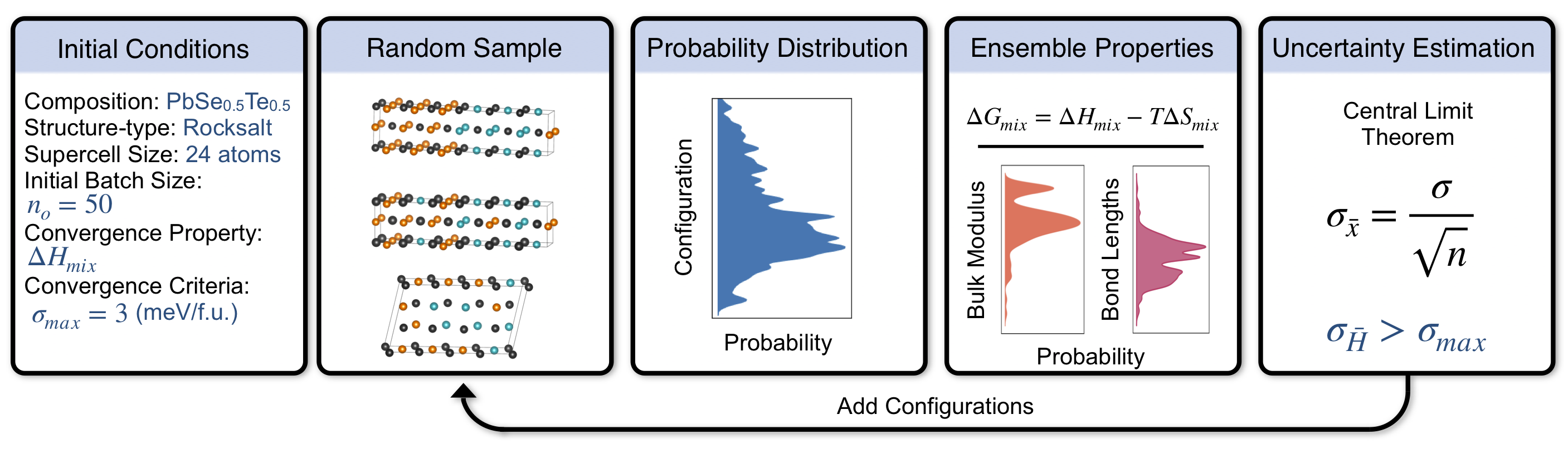}
\caption{\label{fig:flow-chart}The methodology developed herein begins with specifying the initial conditions; random configurations are then generated within those constraints. The $\Delta$H$_{mix,i}$ and the temperature-dependent probability of each configuration are calculated. The properties of interest are then determined using ensemble averaging. Additionally, $\Delta$G$_{mix}$ is calculated to determine stability with respect to the free energy convex hull. The central limit theorem is then applied to estimate the uncertainty in the predictions ($\sigma_{\bar{X}}$) and the number of new configurations that may be needed to further reduce the uncertainty below the convergenve criteria, $\sigma_{max}$.}
\end{figure*} 

\section{\label{sec:level1}Methods}
An overview of the presented method is shown in Fig.~\ref{fig:flow-chart}. First, the initial conditions are established. Alloy configurations are then randomly sampled from the set of total of configurations. Total-energy calculations are run on the configurations, plus the additional calculations that are needed to compute the properties of interest. Then, the temperature-dependent probabilities of these configurations are determined using statistical mechanics. The ensemble properties are derived from the probabilities of the configurations and their properties. For the property of interest, the uncertainty in the prediction is estimated using the Central Limit Theorem. If the uncertainty is above the specified convergence criteria, the Central Limit Theorem is used to determine the number of additional configurations that are needed. Finally, these calculations are conducted, and the ensemble property is reevaluated. 

\subsection{\label{sec:strucgen}Configuration Sampling}
As referenced in the first panel of Fig.~\ref{fig:flow-chart}, the desired composition, structure-type, supercell size, and number of initial configurations are specified. The convergence property and criteria need to be specified as well, but we will discuss this in \ref{sec:CLT3}. For a given supercell size, a number of symmetry-inequivalent supercells are constructed. This is done by using the algorithm developed by Hart and Forcade \cite{hart2008algorithm}, which produces Hermite Normal Form transformation matrices that correspond to symmetry-inequivalent supercells. A complete list of possible decorations is then built for each supercell. Here, a configuration is defined as the pairing of a supercell and a decoration. The size of the complete set of configurations is simply the product of the number of supercells and decorations. A randomly chosen configuration is one that is sampled from this set. 

All transformation matrices used provide symmetry inequivalent supercells, but once decorations are applied, some configurations will be symmetrically equivalent. Both translational and rotationally symmetric configurations are included in the ensemble. They are interpreted to be degenerate, and thus should be counted within the ensemble, as is done elsewhere \cite{aflowpocc}. Finally, the chosen configurations are subjected to full structural relaxations, which was done using VASP \cite{kresse_CMS:1996}. The generation of configurations was executed in Python using the Pylada software \cite{d2010pylada}. The numerical approaches used this work are described in detail in Section \ref{sec:Comp_Details}.

\subsection{\label{sec:Ensemble_Avg}Probability Distribution and Ensemble Properties}

\begin{figure}[t!]
\includegraphics[width = 0.9 \linewidth]{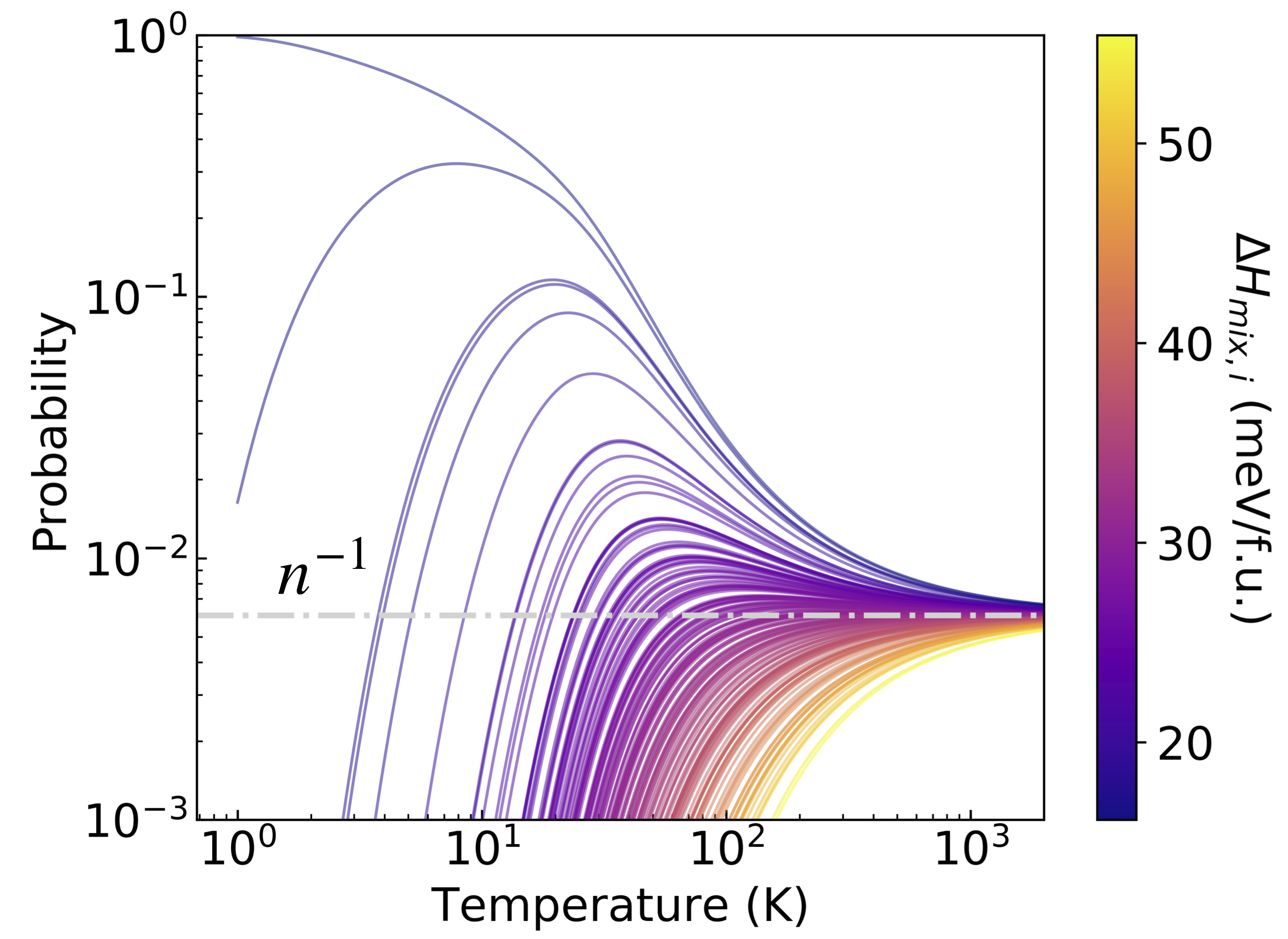}
\caption{\label{fig:prob} Within a given ensemble, the variation in $\Delta H_{mix,i}$ leads to significant 
temperature evolution of the probability for each configuration. 
 In the high-temperature limit,   the probabilities converge to $n^{-1}$.  
 Here, an ensemble of 50 PbSe$_{0.5}$Te$_{0.5}$ configurations is considered, with each configuration represented by a line colored by its $\Delta H_{mix,i}$.} 
\end{figure}

The first step is to calculate the partition function, $Z$:
\begin{equation}\label{parition function}
    Z=\sum_i^n e^{-E_i/k_{B}T},
\end{equation} 
where $i$ is the configuration index, $n$ is the total number of configurations sampled, $E_{i}$ is the total energy per formula unit of a given configuration, $k_B$ is Boltzman's constant and $T$ is the absolute temperature. The ensemble probability of the $i$th configuration, $P_i$ is
\begin{equation}
    P_i = \frac{e^{-E_i/k_{B}T}}{Z}.
\end{equation}
Illustrated in Fig.~\ref{fig:prob} are the ensemble probabilities of 50 configurations of PbSe$_{0.5}$Te$_{0.5}$. The probability of the ground state configuration approaches unity as temperature decreases, while in the high-temperature limit, the probability of each configuration asymptotes to $n^{-1}$. Generally, the ensemble average of a property, $\tilde{X}_n$, is given by taking the weighted average over $n$ configurations,
\begin{equation}
    \tilde{X}=\sum_i^n P_iX_i.
\end{equation}
Depending on the property, ensemble averaging may require more nuanced averaging--for example, the ensemble radial distribution function in \cite{jones2020glassy}. 

Importantly, when the entropy or free energy of mixing are mentioned, they refer to the configurational free energy and entropy. Vibrational degrees of freedom are not considered here. As Esters et al.~show, vibrational entropy should be considered when the parent compounds have different nearest neighbor environments from the resulting alloy \cite{esters2021settling}. This is not the case in our work, since we consider rocksalt alloys and parent compounds that are either rocksalt, or distorted versions of rocksalt. The exclusion of the vibrational degrees of freedom may affect the accuracy of the mixing temperatures. 

The configurational free energy of mixing, $\Delta G_{mix}$, can be calculated from the partition function in a standard way:
\begin{equation}
     \Delta G_{mix}=-k_B T ln(Z)-\sum_{j}^K E_{j}x_{j}.
\end{equation}
This is value will need to be adjusted once the $\Delta S_{mix}$ correction is applied. In this work, we will be referring to pure constituents that make up the alloy as parent compounds, but the framework presented here would also apply if only elements were being alloyed. Here, $j$ counts over parent compounds of the alloy, $x_j$ is the fraction of the alloy made up by that parent, and $E_{j}$ is the total energy of the parent. Finally, $K$ is the number of parents that comprise the alloy.
The enthalpy of mixing, $\Delta H_{mix,i}$, for configuration $i$, is defined in the following equation:
\begin{equation}\label{eq:Emix}
\Delta H_{mix,i}=(E_{i}+pV_i)-\sum_{j}^K x_{j}(E_{j} + pV_{j}).
\end{equation} \par
Here, $p$ is the pressure and $V_i$ and $V_j$ are the volumes per formula unit. All results presented here correspond to the low pressure ($p\approx{}0$) case for which  internal energy and enthalpy are equal, as are the Helmholts and Gibbs free energies. The formalism can easily be extended to elevated pressures by computing the equations of state, E(V), for both the randomly sampled configuration and their parent compounds.  
 
$\Delta H_{mix}$ for the ensemble is then defined by ensemble averaging over all $\Delta H_{mix,i}$. Next, the entropy of mixing, $S_{mix}$, can be calculated from the ensemble averaged enthalpy of mixing $\Delta H_{mix}$ and the free energy of mixing:
\begin{align}\label{eq:s-numeric}
    \Delta S_{mix}(T)=\frac{\Delta H_{mix}(T)-\Delta G_{mix}(T)}{T}.
\end{align}

$\Delta S_{mix}$, $\Delta G_{mix}$, and to a lesser extent, $\Delta H_{mix}$, are dependent on the number of configurations sampled. Random sampling approximates well the distribution of energies, and hence, its ensemble average. Thus, $\Delta H_{mix}$ will remain relatively constant after convergence with respect to the number of configurations has been achieved. However, $\Delta S_{mix}$, and consequently $\Delta G_{mix}$, will change because the partition function increases monotonically with the number of configurations.

To remove the dependency of $\Delta S_{mix}$ on the number of randomly sampled configurations, a scaling factor is applied to $\Delta S_{mix}$ so that it always asymptotes to the configurational entropy in the high-temperature limit, which is analytically known. This value, which we will call $\Delta S_{\infty}$, can be calculated using the standard formula for the configurational entropy of fully random alloys:
\begin{equation}
    \Delta S_{\infty}=-k_{B}\sum_{i} x_i  ln(x_i),
\end{equation}
where ${i}$ is a parent compound of the alloy and $x_i$ is its alloy fraction. A scaling factor, $d$, is defined as:
\begin{equation}
	\frac{\Delta S_{\infty}}{\Delta S_{mix}(T\rightarrow{}\infty)}=d.
\end{equation}
 In this work, $T=2,000$\,K is sufficiently high such that the entropy has asymptoted. This temperature was chosen based on the inspection of the systems present in this paper, but it may need to be reconsidered for other systems. The scaling factor is then applied to the entire $\Delta S_{mix}$ curve. From there, the free energy of mixing can then be recalculated using this scaled final entropy of mixing.

\subsection{\label{sec:CLT3}The Central Limit Theorem and Supercell Size}
\begin{figure*}
\centering
\includegraphics[width=0.75\linewidth]{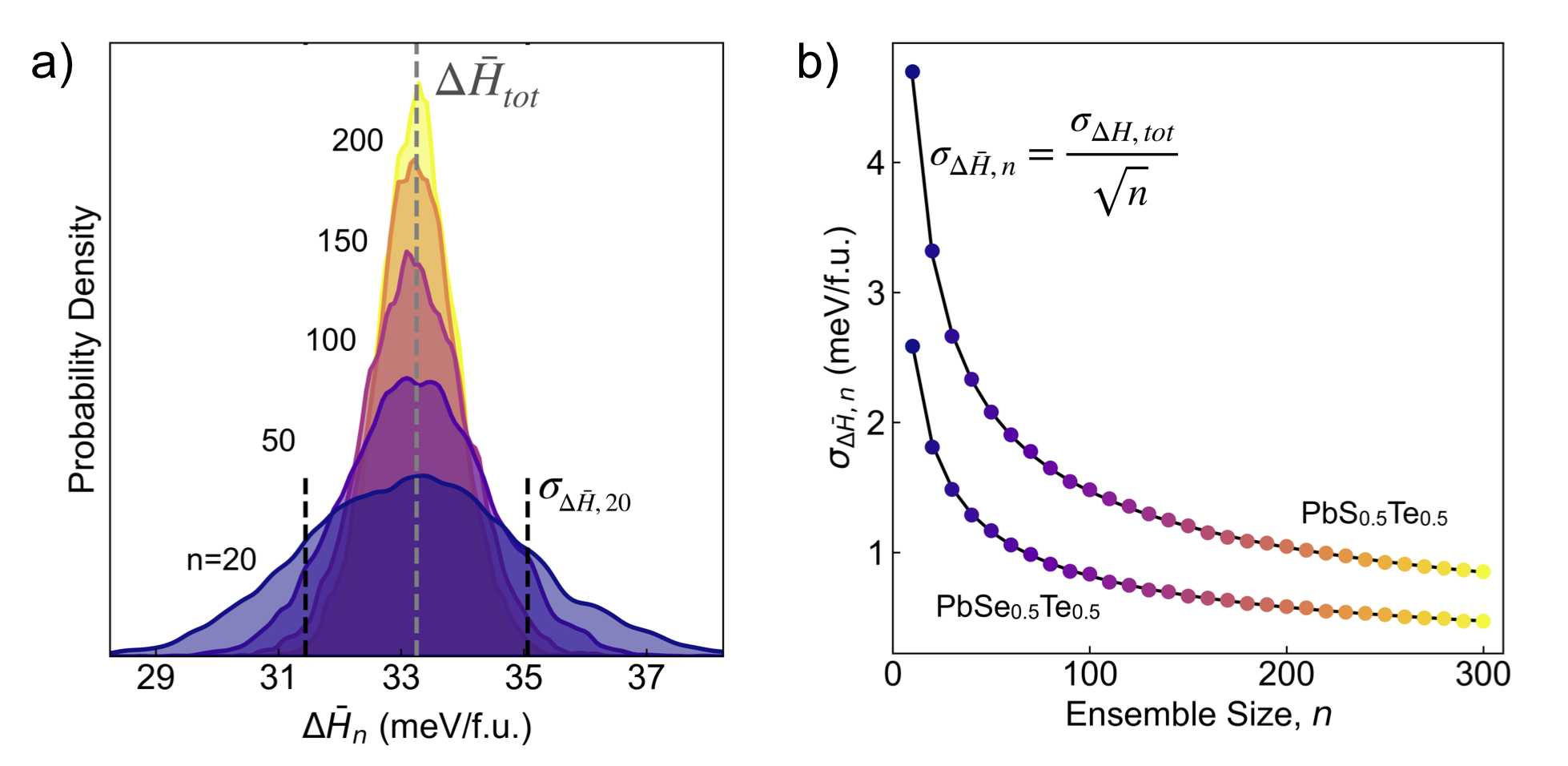}
\caption{\label{fig:clt} To illustrate the CLT, we start with the composition PbSe$_{0.5}$Te$_{0.5}$ and highlight in (a) that the distribution of $\Delta{}\bar{H}_n$ values for any given $n$ forms a Gaussian centered around the true value, $\Delta{}\bar{H}_{tot}$. The standard deviation of each Gaussian, $\sigma_{\Delta{}\bar{H},n}$, decreases with increasing $n$. Each Gaussian arises from calculating $\Delta{}\bar{H}_n$ of 10,000 ensembles.  Each point in (b) is determined by taking the standard deviation in $\Delta{}\bar{H}$ from 10,000 ensembles, including the Gaussians in (a). In (b), this decrease in $\sigma_{\Delta{}\bar{H},n}$ can be modeled using the central limit theorem (black line). By simply knowing the ensemble size and $\sigma_{\Delta{}H,tot}$, $\sigma_{\Delta{}\bar{H},n}$ can be determined. The procedure is repeated for PbS$_{0.5}$Te$_{0.5}$ to illustrate the importance of chemical composition. Since $\sigma_{H,tot}$ is larger, PbS$_{0.5}$Te$_{0.5}$ requires more configurations to reach the same level of uncertainty.}
\end{figure*}
When faced with an overwhelmingly large configurational space, random sampling is one approach to render the number of calculations tractable. Random sampling can be viewed as an acceptable approximation under the following condition. For a property of interest, $X$, the ensemble average derived from $n$ randomly sampled configurations, $\tilde{X}_n$, is in agreement with the ensemble average of the total distribution of configurations, $\tilde{X}_{tot}$ (i.e.~$\tilde{X}_n\approx{}\tilde{X}_{tot}$). In practice, however, $X_{tot}$ will not be available. Therefore, an alternative method for verifying random sampling is required. 

The Central Limit Theorem (CLT) offers one such path \cite{CLT,CLT-History}. It states that if the above random sampling procedure were repeated infinitely many times, the resulting distribution of $\tilde{X}_n$ values would have three main characteristics. It would be i) approximately a Gaussian distribution that is ii) centered around the true mean, $\tilde{X}_{tot}$, and iii) the distribution would have a standard deviation of $\sigma_{\tilde{X},n}$. Critically, $\sigma_{\tilde{X},n}$ is the uncertainty in our $X_n$ prediction due to random sampling. Evaluating $\sigma_{\tilde{X},n}$ will be crucial for justifying the random sampling implemented in this work. 

In a simpler form, the CLT applies for sampling from the true distribution. In our case, we are sampling configurations randomly from a uniform distribution, and thus, the simpler form of the CLT applies for when all configuration probabilities are equal. This happens in the high-temperature limit. The uncertainty in the ensemble averages are thus for the high-temperature limit.

The CLT allows us to do this; in the limit as $n\rightarrow{}\infty$, the CLT states:
\begin{equation}\label{eq:CLT}
    \sigma_{\bar{X},n}=\frac{\sigma_{X,tot}}{\sqrt{n}}.
\end{equation}

For finite $n$, eq.~\eqref{eq:CLT} is an approximation. However, the left side of eq.~\eqref{eq:CLT} asymptotes to the right side for fairly small $n$. Often, an $n$ of 30 is deemed sufficient such that eq.~\eqref{eq:CLT} holds \cite{islam2018sample,mendez1991understanding}. As will be shown in Fig.~\ref{fig:clt}, our numerical simulations for the systems in this work are in general agreement; we illustrate that using an $n>=20$ results in $\bar{X}$ distributions that are approximately Gaussian with standard deviations that agree with eq.~\eqref{eq:CLT}.

We will now be moving on to the second approximation within the CLT. Given a particular ensemble of $n$ randomly chosen configurations where property $X$ has been calculated for each configuration, the resulting distribution of $X$ values has a standard deviation of $\sigma_{X,n}$. When $n$ is sufficiently large, $\sigma_{X,n}$ is approximately equal to $\sigma_{X,tot}$, as will be shown in the Results. One can then make the following approximation:
\begin{equation}\label{eq:CLT2}
    \sigma_{X,n} \approx{} \sigma_{X,tot}.
\end{equation}
In this way, the uncertainty of $\bar{X}$ can be estimated from one sufficiently large ensemble of configurations. 

In practice, a modestly large ensemble of initial size $n_o$ is built and the above approximation is made. If the uncertainty in $\bar{X}$ is larger than desired, one can rearrange eq.~\eqref{eq:CLT} using the desired $\sigma_{\bar{X},n}$ and $\sigma_{X,n_o}$ to estimate the additional calculations needed to achieve the desired uncertainty. After the additional calculations are completed, one can reassess the uncertainty using eq.~\eqref{eq:CLT}. Overall, the described approach  efficiently establishes $\bar{X}$ within the desired uncertainty, while keeping the number of calculations to a minimum.  

In applying the CLT to the independent cell approximation, we will need to determine an appropriate $n_o$ that is sufficiently large to estimate $\sigma_{X,tot}$ while remaining computationally efficient. In practice, one needs to perform a proper convergence test to find a suitable $n_o$. In our work, we determine the appropriate $n_o$ in a different way by using an exhaustive enumeration of all configurations, which is done to illustrate the methodology and compare our findings with exact results. 

A significant portion of the Results will be centered around the thermodynamics of mixing. We therefore chose to illustrate CLT with $\Delta H_{mix}$. Here we use the high-temperature limit of $\Delta H_{mix}$ ($\Delta H_{mix}(T\rightarrow{}\infty)=\Delta{}\bar{H}$), such that $\Delta{}\bar{H}$ is the simple average of $\Delta H_{mix,i}$ across all sampled configurations. We use the $\Delta H_{mix,i}$ of all $\sim$37,000 24-atom configurations of PbSe$_{0.5}$Te$_{0.5}$ to make our total distribution. From the total distribution, 10,000 different ensembles of $n$ configurations are randomly generated and the corresponding 10,000 $\Delta{}\bar{H}_n$ values are calculated. Furthermore, as shown in eq.~\eqref{eq:CLT}, the standard deviation in the distribution of the 10,000 $\Delta{}\bar{H}_n$, $\sigma_{\Delta{}\bar{H},n}$, changes as a function of $n$. To illustrate this effect, we performed the above procedure, building 10,000 ensembles for various $n$, ranging from 20 to 300. Although we use the high-temperature limit for simplicity of illustration, the CLT can be applied for any temperature; the standard deviation must reflect that not all configurations have equal probabilities. As spoken about in the Discussion, doing so requires further caution around sampling.

Fig.~\ref{fig:clt}a shows the associated distributions of $\Delta{}\bar{H}_n$ values for various $n$. All of the distributions form an approximate Gaussian, as expected from the CLT. If the number of ensembles were infinite instead of 10,000, the distributions would truly be Gaussians. As $n$ increases from 20 samples per ensemble to 300, it is apparent that $\sigma_{\Delta{}\bar{H},n}$ decreases. In other words, if one were to randomly sample 20 configurations of PbSe$_{0.5}$Te$_{0.5}$, there is a large range of probable $\Delta{}\bar{H}_{20}$ values that could occur, while sampling 200 configurations would significantly narrow this range.

Fig.~\ref{fig:clt}b further illustrates how the size of $n$ reduces uncertainty in the calculation of $\Delta{}\bar{H}_n$. The standard deviation of each Gaussian, $\sigma_{\Delta{}\bar{H},n}$, in panel \textit{a} is plotted, as well as for many other $n$ that were not shown in panel \textit{a}. For comparison, the same procedure was repeated for PbS$_{0.5}$Te$_{0.5}$, which has a larger $\sigma_{\Delta{H},tot}$. The two lines plotted in Fig.~\ref{fig:clt}b are generated from eq.~\eqref{eq:CLT}.  

Composition affects uncertainty as well. The uncertainty in $\Delta{}\bar{H}$ is smaller for PbSe$_{0.5}$Te$_{0.5}$ than PbS$_{0.5}$Te$_{0.5}$, reflecting the wider distribution of configurations in the latter. For a given uncertainty criteria, PbS$_{0.5}$Te$_{0.5}$ requires more samples. However, for $\Delta{}\bar{H}$ in the studied systems, the uncertainty is kept below 3 meV/f.u. across the entire composition space using just 50-55 configurations. We employ the CLT throughout the work. In Fig. \ref{fig:clt2}, we will return to the CLT and illustrate that $\sigma_{n} \approx{} \sigma_{tot}$ is a reasonable approximation to make within the context of our method. \par

\textbf{Supercell Size.} Using a sufficiently large supercell size is crucial for calculating the ensemble properties of an alloy. The ensemble statistics can widely vary depending on the supercell size chosen, as will be shown in Fig.~\ref{fig:converge}. The following procedure was adopted for determining the appropriate supercell size. First, evaluate the properties of interest using an SQS for the specified structure-type and composition. It is important to make sure the SQS properly sized as well. Second, for the same composition and structure-type, run a series of randomly sampled configurations. Using the CLT, iteratively add samples until the high-temperature ensemble average result is sufficiently converged. This high-temperature ensemble average should match up well with SQS. If it does not, then repeat the same procedure with configurations of a larger supercell size until the high-temperature ensemble average and the SQS values are within the needed uncertainty.

\subsection{\label{sec:Comp_Details}Computational Details}
Structural relaxations and total energy calculations are conducted within VASP \cite{kresse_CMS:1996}, using the PBE functional \cite{perdew_PRL:1996} within the projector-augmented wave method \cite{bloechl_PRB:1994}. All structural degrees of freedom were allowed to be optimized within the structural relaxation (ie. volume, cellshape, atom positions). A planewave cutoff of 340 eV and a gamma-centered k-point mesh are used such that the energy is converged to within 3 meV/atom.  

We use the AFLOW-POCC method \cite{aflowpocc} for conducting complete sampling for PbSe$_{0.5}$Te$_{0.5}$ and PbS$_{0.5}$Te$_{0.5}$. The SQS structure was generated using Monte Carlo SQS (mcsqs) within the ATAT framework \cite{mcsqs}. Clusters were randomized up to the sixth nearest neighbors for pairwise interactions, and first nearest neighbors for three-wise interactions. The bulk modulus was calculated by fitting the Birch–Murnaghan equation of state. For a given structure, this involved generating multiple volumes near the minimum volume of the structure and calculating the energy of those volumetrically scaled structrues. The equation of state was then fit from those total energy calculations.\cite{birch1947finite,murnaghan1944compressibility}. The ensemble bulk modulus was calculated by taking the Reuss average over all configurations. The reuss average was used since it assumes that all configurations are under equivalent stress \cite{reuss1929berechnung}. For PbSe$_{0.5}$Te$_{0.5}$, the bulk modulus was calculated for all configurations with supercell sizes up to 20 atoms. For 24-atom supercells, the bulk modulus was calculated for 200 randomly sampled configurations instead of all 1107 non-degenerate configurations in order to keep the test computationally affordable. 

Determining the coordination number of an atom inherently requires making a somewhat arbitrary decision about what atoms are considered to be nearest neighbors. Our method is the following: for a given atom, we find the closest neighboring atom, set that as our base bond length, and only include other atoms into the first shell of coordination if their distance from the central atom is within 20$\%$ of the base bond length. We find that a tolerance of 20$\%$ helps to capture distorted bonding environments while excluding the second coordination shell. 

\section{\label{sec:Results}Results}
We will be working through multiple case examples to validate this method and illustrate its various applications. To start, we will study a single composition, PbSe$_{0.5}$Te$_{0.5}$, before moving on to the complete PbSe$_{1-x}$Te$_{x}$ pseudo-binary where we will derive the free energy of mixing across the composition space, all as a function of temperature. Subsequently, the Pb-chalcogenide pseudo-ternary (PbS$_{1-x-y}$Se$_{x}$Te$_{y}$) will be explored to show broader trends and the use of the free energy convex hull as it compares to experiment. Finally, a high-entropy system, (Pb$_{1-x}$Ge$_{x}$)(Te$_{y}$Se$_{z}$S$_{1-y-z}$), will be studied.  

\subsection{\label{sec:Supercell Size} PbSe$_{0.5}$Te$_{0.5}$}
For the single composition PbSe$_{0.5}$Te$_{0.5}$, we will be focusing primarily on the thermodynamics of mixing. To show that the results are sufficiently converged with respect to random sampling, the Central Limit Theorem (CLT) will be used. We take this opportunity to illustrate the main assumptions of the CLT within the context of this method. Next, the supercell size necessary to capture acruately the  thermodynamics of mixing needs to be determined; we thus calculate the high-temperature ensemble averages for varying supercell sizes and compare them to the SQS value. Finally, we briefly illustrate the dependence of ensemble supercell size on the bulk modulus. This is done to illustrate the importance of moving to larger supercells when considering ensemble properties beyond thermodynamics. 

\begin{figure}[t]
\includegraphics[width = \linewidth]{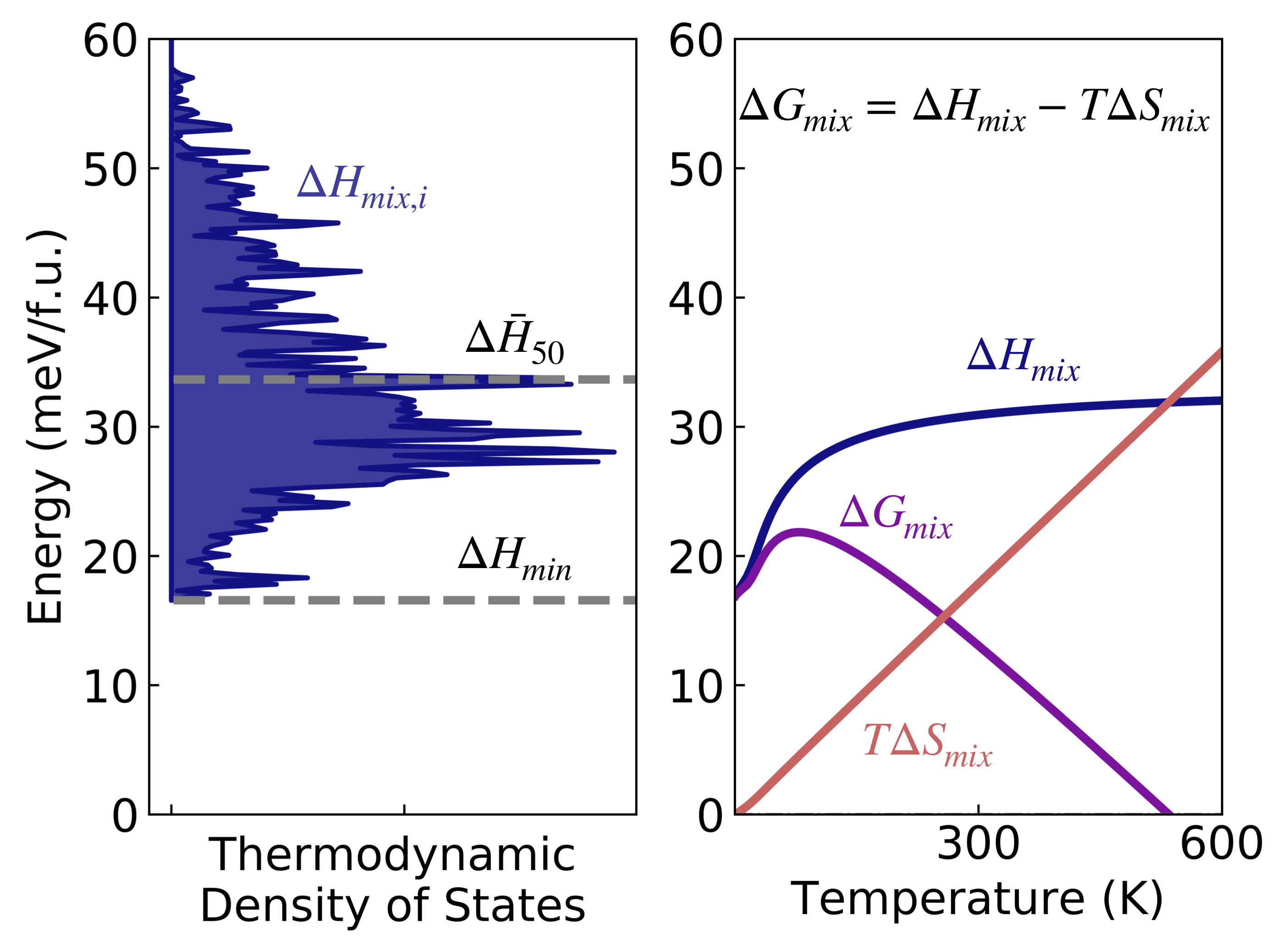}
\caption{\label{fig:standard} The distribution of $\Delta H_{mix,i}$ is shown in the left panel, made from an ensemble of 50 PbSe$_{0.5}$Te$_{0.5}$ configurations. $\Delta \bar{H}_{50}$ denotes the average $\Delta H_{mix,i}$ and $\Delta H_{min}$ shows the ground state configuration. In the right panel, the ensemble $\Delta H_{mix}$ approaches $\Delta \bar{H}_{50}$ as the temperature increases. The entropic contributions ($T\Delta S_{mix}$) and the resulting free energy $\Delta G_{mix}$ are also shown. For temperatures where $\Delta G_{mix}$ is below zero, the alloy is stable against decomposition to its parent compounds (i.e., PbSe and PbTe).}
\end{figure}

\textbf{Thermodynamics of Mixing.} The thermodynamics of mixing for PbSe$_{0.5}$Te$_{0.5}$ was calculated according to the equations in section \ref{sec:Ensemble_Avg}. In Fig.\ \ref{fig:standard}, the Thermodynamic Density of States (TDOS) in the left panel shows the distribution of states resulting from 50 randomly sampled configurations. The resulting ensemble $\Delta H_{mix}$, $\Delta S_{mix}$, and $\Delta G_{mix}$ are shown in the right panel. At 0\,K, $\Delta H_{mix}$ is equal to the ground state configuration at that composition. As the synthesis temperature increases, $\Delta H_{mix}$ rises, approaching the average of the TDOS. The $\Delta S_{mix}$ is convex at low temperatures, and in this particular system, it asymptotes to its high-temperature limit above ~50\,K, resulting in an ostensibly linear $T\Delta S_{mix}$ curve. The $\Delta G_{mix}$ becomes negative at $\approx{}$550\,K, and thus will not decompose into its parent compounds after that temperature. However, a negative $\Delta G_{mix}$ does not necessarily imply that the composition will be stable. Later sections will show the importance of using the free energy convex hull in determining stability.

\textbf{Determining Sufficient $n_o$ for CLT.} As stated in the Methods, a convergence test must be conducted in order to show that the number of randomly sampled configurations in the initial ensemble, $n_o$, is sufficiently large to satisfy eq.~\eqref{eq:CLT2}. Here, we run a more advanced convergence test by assembling statistics on $\sigma_{\Delta H,n}$ values on 10,000 unique ensembles of PbSe$_{0.5}$Te$_{0.5}$ and PbS$_{0.5}$Te$_{0.5}$. \par

\begin{figure}
\includegraphics[width=0.8\linewidth]{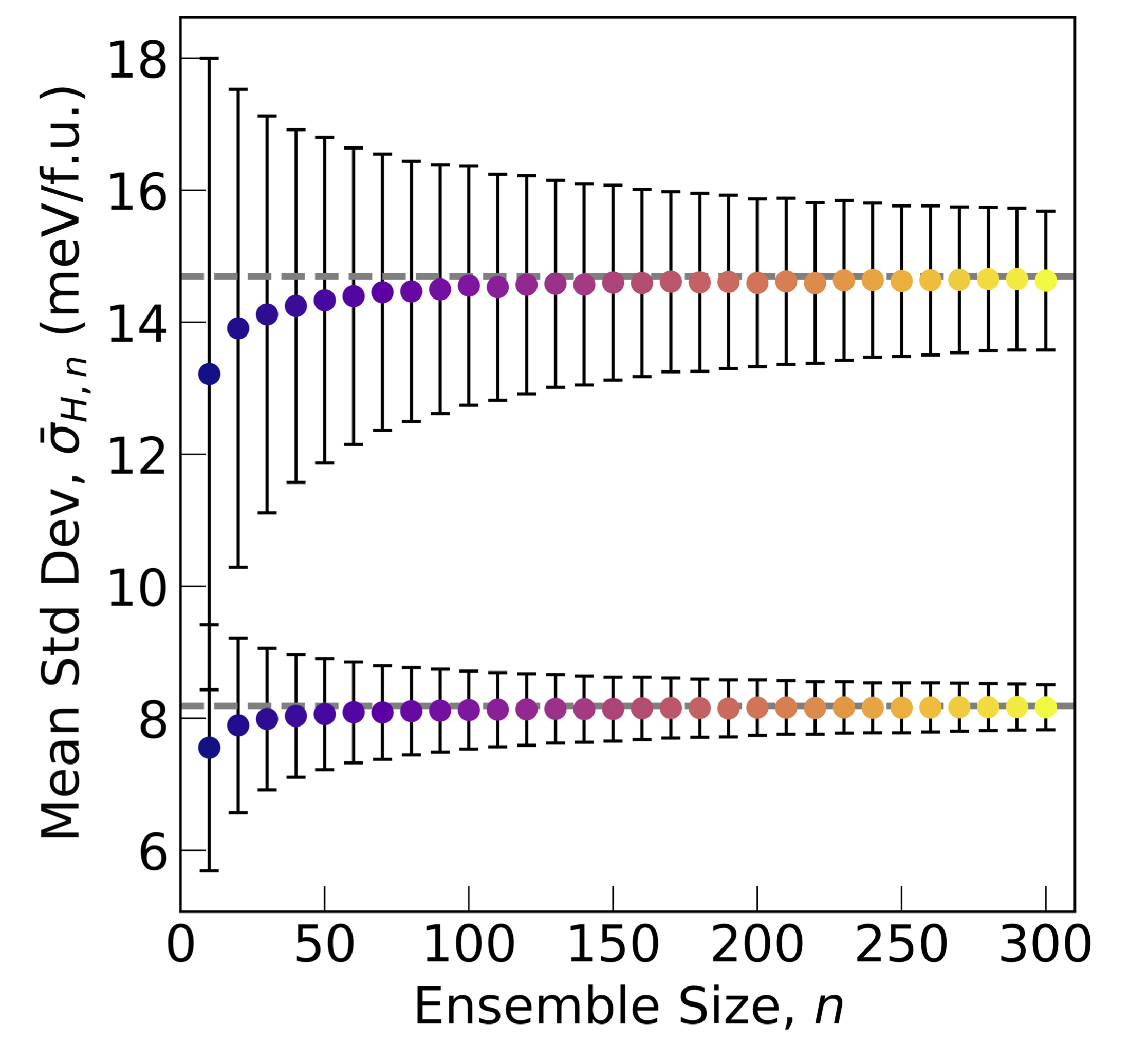}
\caption{\label{fig:clt2}Given the complete ensemble, there is a standard deviation in the values of $\Delta H_{mix}$ ($\sigma_{H,tot}$) denoted by the horizontal grey dashed lines.  
With increasing ensemble size, the mean standard deviation $\bar{\sigma}_{H,n}$ rapidly approaches $\sigma_{H,tot}$.
However, the $\sigma_{H,n}$ of a single ensemble shows greater variation as denoted by the error bars, corresponding to $\pm$ one standard deviation from the mean. The small variation from $\sigma_{H,tot}$ highlights the accuracy of the approximation of Eq. \eqref{eq:CLT2}. 
The larger variation in $\sigma_{H,n}$ for PbS$_{0.5}$Te$_{0.5}$ arises from a broader distribution of $\Delta H_{mix}$ values in the complete ensemble.}
\end{figure}

Fig.~\ref{fig:clt2} assesses how $\bar{\sigma}_{\Delta H,n}$ changes with increasing ensemble size and compares this result to $\sigma_{\Delta H,tot}$, derived from all configurations within the complete distribution. As $n$ increases, $\bar{\sigma}_{\Delta H,n}$ quickly asymptotes to $\sigma_{\Delta H,tot}$. The error bars represent one standard deviation from $\bar{\sigma}_{\Delta H,n}$. Further, the error bars shrink, corresponding to a decrease in the uncertainty of $\sigma_{\Delta H,n}$ for a given ensemble. For $n$ of 50 configurations and above, the uncertainty in the estimates of $\sigma_{\Delta H,tot}$ are fairly small. As the uncertainty decays, we can increasingly justify the assumption of Eq.~\eqref{eq:CLT2}. 
To actually assess the uncertainty contribution from making the approximation in Eq.~\eqref{eq:CLT2}, the error bars in Fig.~\ref{fig:clt2} can be divided by $\sqrt{n}$. 
For instance, consider an $n=50$ ensemble of PbS$_{0.5}$Te$_{0.5}$. On average, the $\sigma_{E,50}$ value is $14 \pm 2$ meV/f.u. Calculating the uncertainty in $\Delta{}\bar{H}$ using Eq.~\eqref{eq:CLT} and Eq.~\eqref{eq:CLT2} results in $\sigma_{\Delta{}\bar{H},50}=2.0 \pm 0.3$ meV/f.u. In this particular instance, we are focused on determining the uncertainty in $\sigma_{\Delta{}\bar{H},50}$, not its actual value. Having an uncertainty in $\sigma_{\Delta{}\bar{H},50}$ of 0.3 due to Eq.~\eqref{eq:CLT2} is insignificant and well within the typical noise of thermodynamic calculations using DFT. We thus conclude that, for our system and property of interest, an initial ensemble size of 50 is sufficiently large such that $\sigma_{H,n} \approx{} \sigma_{H,tot}$. In the remainder of the results, we will adopt an $n_o$ of 50 as a starting point for estimating $\sigma_{\bar{H},n_o}$. As detailed in the methods, additional configurations can then be added to the ensemble until $\sigma_{\bar{H},n'}$ is within the desired uncertainty.  

\begin{figure*}[!t]
\includegraphics[width =1\linewidth]{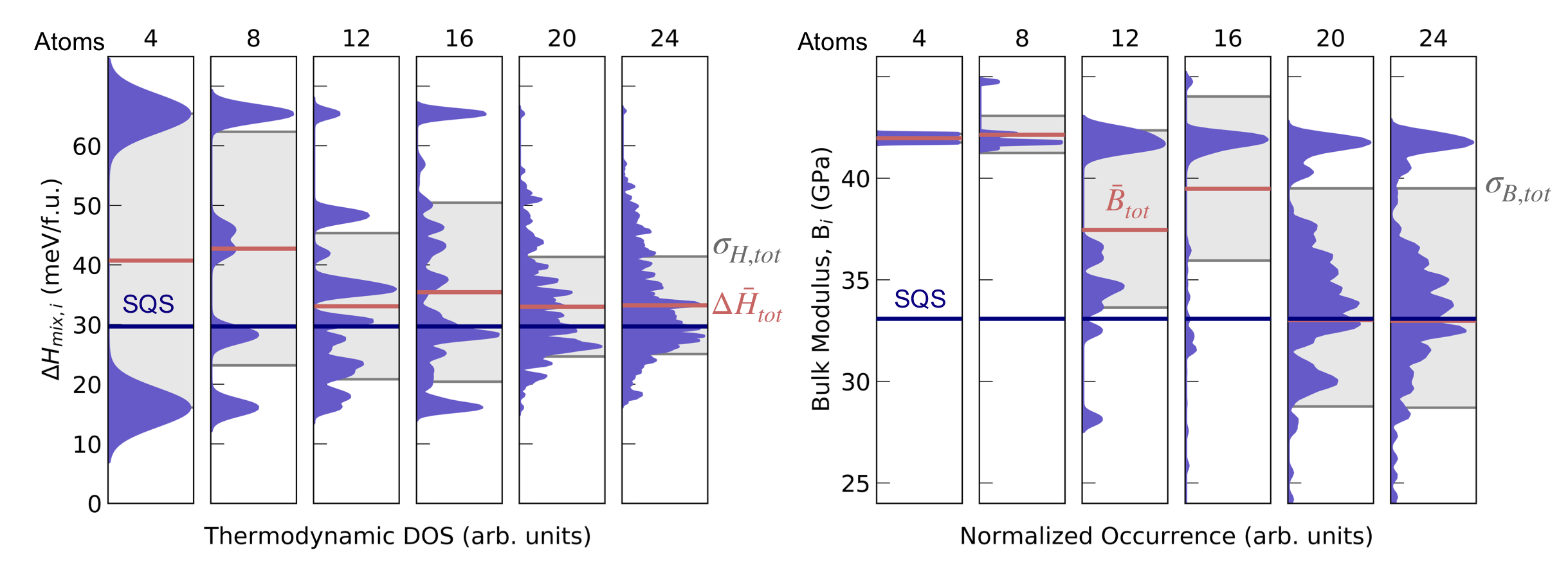}
\caption{\label{fig:converge} (a) To assess the minimum required supercell size, we increase the supercell size for PbSe$_{0.5}$Te$_{0.5}$ until the distribution converges and the $\Delta \bar{H}_{tot}$ of the distribution (red) has come into agreement with the $\Delta H_{mix}$ of a 128-atom SQS supercell (blue). The convergence of the distributions can also be tracked via the $\sigma_{\Delta{}H,tot}$ values (grey). Here, each distribution is built from complete sampling of PbSe$_{0.5}$Te$_{0.5}$ for a given supercell size. (b) The importance of using larger supercells is even more visibly important for a material property like the bulk modulus. For supercell sizes of 12 and 16 atoms, the distributions are significantly different from both the properly converged 20 and 24-atom distributions and the SQS-128.}
\end{figure*}

\textbf{Supercell Size Convergence.} To disentangle the effects of random sampling from supercell size dependence, we ran our convergence tests by sampling all possible configurations from 4 to 24-atom supercells. The results were also compared to a large (128-atom) SQS of the same composition. Since the SQS represents the ensemble in the high-temperature limit, the SQS energy should be equivalent to the mean energy of the thermodynamic density of states, $\Delta{}\bar{H}$.

In Fig.~\ref{fig:converge}, $\Delta{}\bar{H}$ begins to converge with the SQS-128 result by 12 atoms per cell, with the difference decreasing to below 6 meV/f.u. However, the distribution of energies does not converge until 20-24 atoms; here complete sampling involves 200 and 1107 unique configurations, respectively. This convergence can be seen in the thermodynamic density of states and their respective standard deviations. Only minor differences are found between the 20 and 24 atom cells; the differences between the mean energy and the SQS-128 are 3.3 and 3.5  meV/f.u., and the standard deviations are 8.1 and 8.3 meV/f.u., respectively. Convergence with respect to supercell size and the SQS indicates that 20-atom supercells are sufficiently large to incorporate the effects of configurational disorder for this system. These results can be viewed in terms of the CLT. If one were to build an ensemble of 50 randomly sampled configurations and calculated  $\Delta{}\bar{H}_{50}$, one would have a 68 \% of being within 1.17 meV/f.u. of $\Delta{}\bar{H}_{tot}$, (grey shaded region in Fig.~\ref{fig:converge}), and a 95 \% chance of being within 2.34 meV/f.u.

Since energy is not our only property of interest, we converged the supercell size with respect to the bulk modulus. We chose bulk modulus as a test case since it is a medium-cost property that is dependent on atomic structure. In Figure \ref{fig:converge}b, the 12-atom supercells are not large enough to properly represent the softening in PbTe$_{0.5}$Se$_{0.5}$ that comes from distortion. However, by 20 and 24 atoms, we see excellent correspondence between the ensemble average and SQS, as well as the distributions between 20 and 24 atoms. 
Considering the distribution for 24-atom cells, we see significant variation in $B$ and thus obtain a $\sigma_{\bar{B}, tot}$ of 5.4 GPa.  From the CLT, to achieve an uncertainty of less than 1 GPa in $B$, we would require an ensemble with $n = 30$.

\begin{figure}
\centering
\includegraphics[width = 0.9 \linewidth]{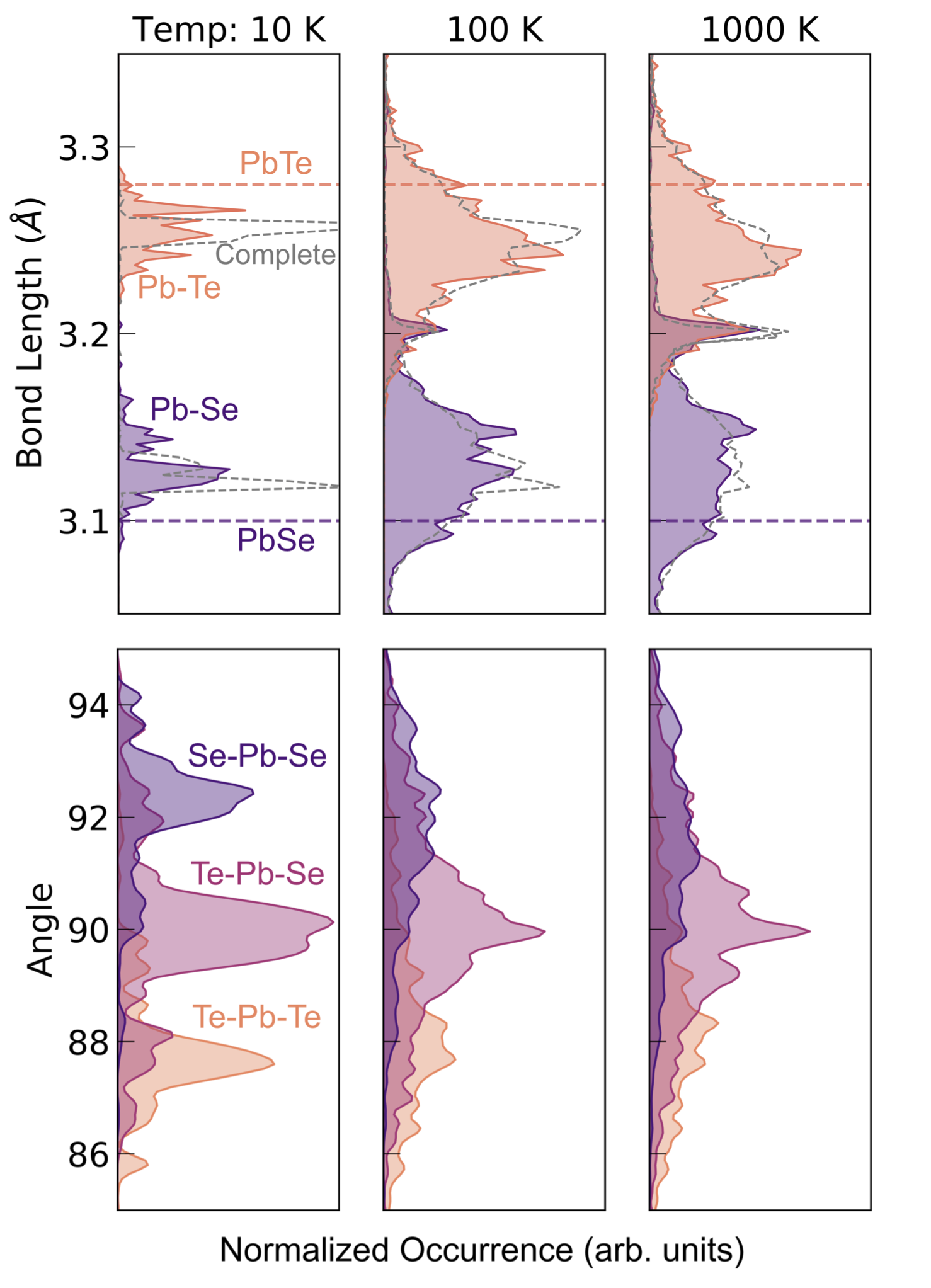}
\caption{\label{fig:Struc-Temp}  At low temperature, PbSe$_{0.5}$Te$_{0.5}$  has distinct Pb-Te and Pb-Se bond lengths and strong clustering of bond angles. 
The deviation from 90$^\circ$ even at low temperature for these rocksalt structures highligths the internal strains present in these alloys. 
With increasing temperature, the distribution of bond lengths and angles broadens significantly.
The x-axis of the 10\,K panel has a scale that is four times larger than the 100\,K and 1,000\,K panels in order to accommodate the large, narrow peaks. Here, the colored distributions are for a random ensemble for 50 configurations and the dashed lines show the complete sampling for bond lengths. Despite the low number of configurations, these structures still involve 3,600 unique bonds and 7,200 angles.}
\end{figure}

\textbf{Local Structure.} By ensemble averaging the structure of relaxed cells, we can explicitly calculate the effects of configurational disorder on short-range structural disorder.  From 50 randomly relaxed cells, there are 3,600 bonds and 7,200 bond angles, originating from a variety of unique local atomic arrangements. The magnitude of data allows for an in-depth statistical analysis of the structure. Furthermore, unlike SQS, the ensemble structure can be calculated as a function of temperature. 

At 10\,K, as shown in Fig.~\ref{fig:Struc-Temp}, the distributions of bond lengths and bond angles have well-defined peaks. In fact, the x-axis for the bond lengths had to be stretched to four times that of 100\,K and 1,000\,K, just to fully include the peaks. The narrowness of the peaks can be, in part, attributed to the configurational probability distribution. At such a low temperature, the four lowest energy configurations make up 95$\%$ percent of the ensemble (as can be seen in Fig.~\ref{fig:prob}). The degree of structural disorder is thus limited by the narrow range of structures present in the ensemble. 

Within the ensemble, Pb-Se bonds are closer to the pure PbSe bond length (shown as a horizontal dashed line), but they are slightly larger due to the presence of Te. The reverse is true for Pb-Te bonds—they are shorter than that of pure PbTe. If the Virtual Crystal Approximation were invoked, and PbSe$_{0.5}$Te$_{0.5}$ were assumed to be perfectly rocksalt, then its single, universal bond length would be 3.19 A, derived from taking the average for that of PbSe and PbTe. There is, however, a significant gap between the two bond length distributions, and ostensibly no amplitude at 3.19 A. The ensemble structure is thus locally distorting from the rocksalt structure, so that the constituent bond lengths may more closely resemble the pure parent compounds. The same has been observed experimentally in ZnSe$_{1-x}$Te$_{x}$ \cite{peterson2001local} and theoretically for PbSe$_{1-x}$Te$_{x}$ \cite{doak2012coherent}. The bond angle distributions at 10\,K are also relatively narrow. Angles that are made up of exclusively Se anions are obtuse to accommodate the larger Te atoms. For angles with Te as the only anion, the angles are acute, and mixed anion angles are centered around 90 degrees, which corresponds to the ideal rocksalt structure. 

As the temperature increases from 10\,K to 100\,K, the probability of higher energy configurations increases as well, resulting in a smearing of the bond length distribution. Interestingly, while the distribution of bond length smears, the average Pb-Te and Pb-Se bond lengths remain constant (within 0.011 and 0.019 {\AA}, respectively). Due to the smearing, there is now amplitude where there was a gap, and there is a slight peak at 3.2 {\AA} that is shared by both distributions, corresponding to the bond length that would be derived from the Virtual Crystal Approximation. The bond angles also significantly smear, but the average still remains at 90 degrees, pointing to a local distortions and the structure being globally rocksalt. For bond lengths at 1000\,K, the peak at 3.2 {\AA} grows, and since there is both Pb-Se and Pb-Te amplitude, it is roughly twice the size of the surrounding peaks. Thus, higher energy structures that adopt a more VCA local structure (i.e.~have bond lengths of 3.19 {\AA}) are being incorporated into the ensemble. The bond angle distributions continue to smear moving from 100\,K to 1,000\,K. 

The same bond length analysis was repeated for the complete set of all ~37,000 configurations, and the resulting distributions are shown as grey dashed lines  in Figure 8. At 10\,K, it can be seen that complete sampling has narrower, larger peaks than random sampling. This is to be expected. Random sampling, which samples from a uniform distribution of configurations, will have difficulty approximating a highly nonuniform distribution where the configurational probabilities vary widely, like one that is seen at 10\,K. Still, the derived probabilities of the various configurations is sufficiently good to show the gap in bond lengths at 10\,K. The average bond lengths at 10\,K for Pb-Te  and Pb-Se are 3.256A and 3.124 for complete sampling and 3.255A, 3.128A for random, showing overall good correspondence as well. Finally, at 100\,K and 1,000\,K, where the probability distribution for configurations is more uniform, random sampling, does an excellent job of replicating the bond length distribution. 

\subsection{\label{sec:PbSe-PbTe} PbSe$_{1-x}$Te$_{x}$ Pseudo-binary}

\begin{figure}[t]
\centering
\includegraphics[width = 0.9 \linewidth]{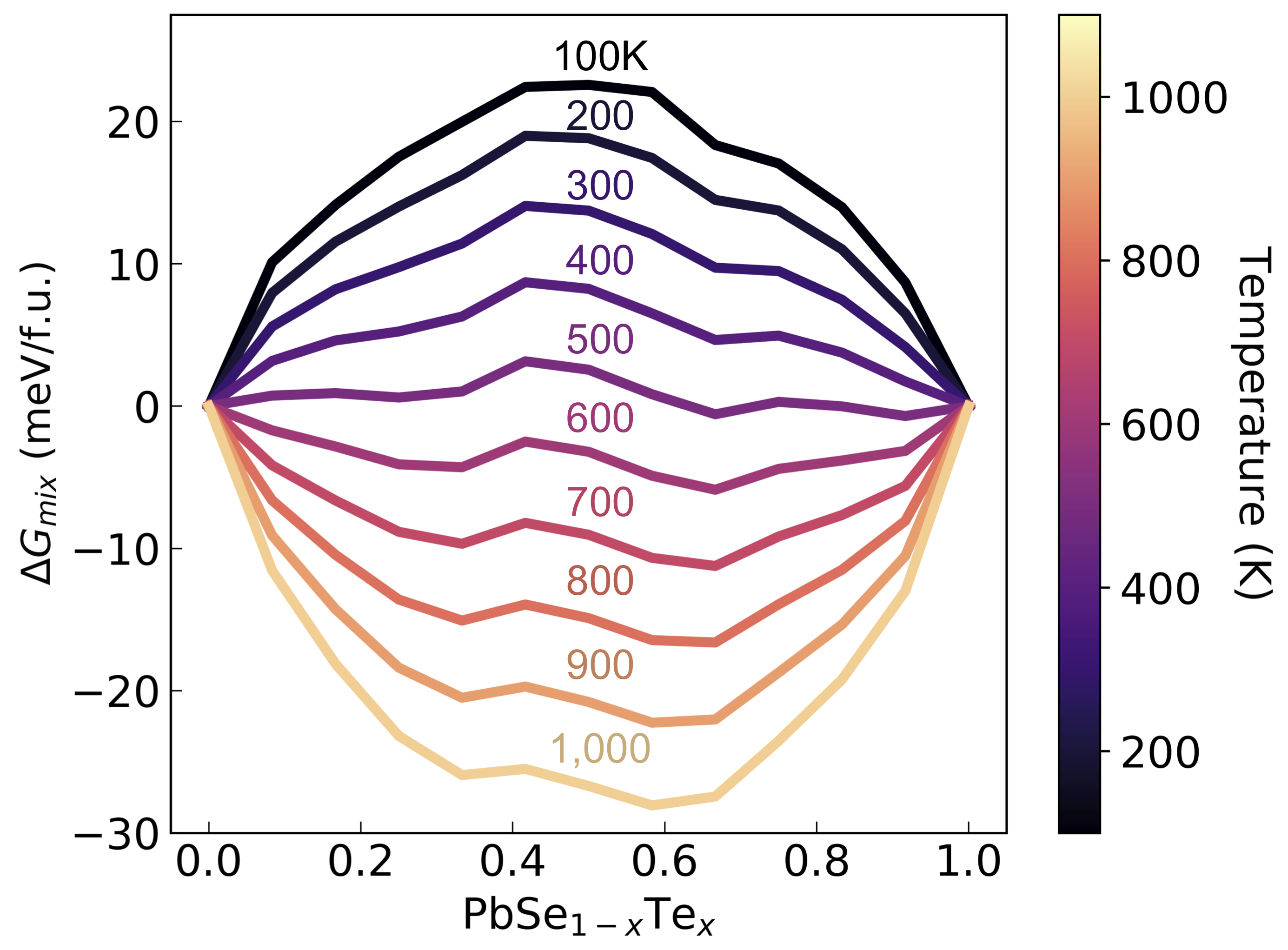}
\caption{\label{fig:free-energy-curve} 
The $\Delta G_{mix}$  of  PbSe$_{1-x}$Te$_{x}$  gradually becomes more concave with increasing temperature, reaching full convexity at 620\,K (within a tolerance of $2\sigma_{\bar{H},n}$). Here, the alloy compositions are sampled at 1/12 increments.
}
\end{figure}
The free energy calculations from Fig.~\ref{fig:standard} were extended across the PbSe$_{1-x}$Te$_{x}$ system. Fig.~\ref{fig:free-energy-curve} plots the $\Delta G_{mix}$ for all thirteen compositions as a function of temperature. The $\Delta G_{mix}$ of every alloy decreases with temperature, but this change is greater for compositions towards the center since they have larger configurational entropies. Thus, the $\Delta G_{mix}$ curve goes from being concave to convex as the temperature increases. At typical growth temperatures of about 770-990\,K, PbSe$_{1-x}$Te$_{x}$ is fully miscible experimentally \cite{ortiz2019towards,PbSSeTe-Phase-Diagram,zhang2012heavy}. The work by Liu et al.~show the temperature at which full miscibility occurs is between 573 and 773\,K. Our results align with these experimental results, showing that PbSe$_{1-x}$Te$_{x}$ is fully miscible at 600\,K, below the typical experimental growth temperatures and within the range that Liu et al report. At elevated temperatures, we find that there are some compositions where the $\Delta G_{mix}$ curve is locally concave, but this deviation from convexity is 1-2 meV/f.u., which is within the uncertainty of random sampling. More specifically, a composition is classified as being on the hull if its $\Delta G_{mix}$ is within 
$2 \times \sigma_{\Delta{}\bar{H},n}$ of the free energy convex hull. 

\begin{figure}[t]
\centering

\includegraphics[width = 0.7 \linewidth]{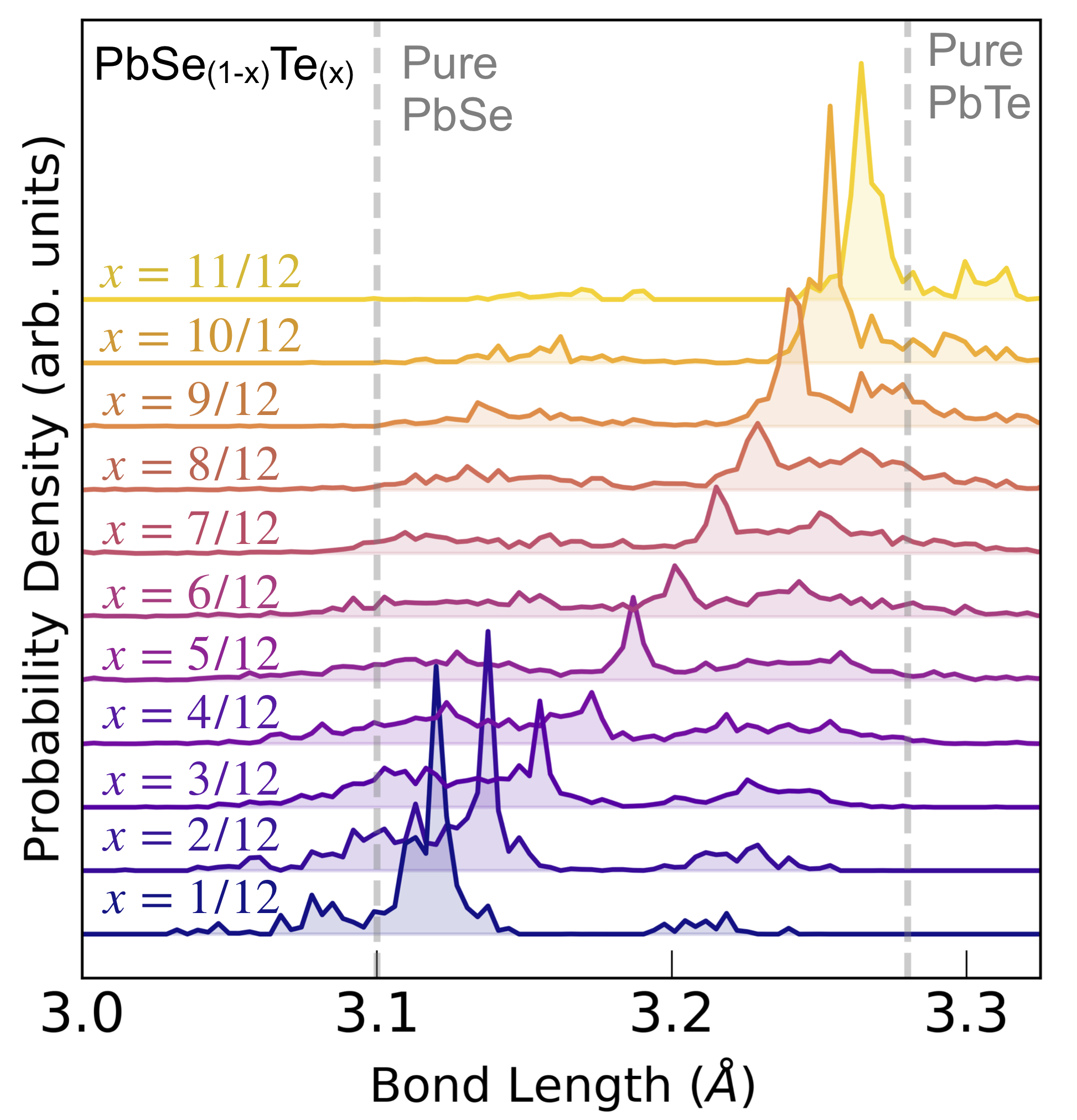}
\caption{\label{fig:pbte-pbse_bonds} Bond length distributions in PbSe$_{1-x}$Te$_{x}$ smear as we move towards x-values of 6/12, highlighting the increased structural disorder in these alloys. The peak of each distribution corresponds well to Vegard's law.
All distributions are shown in the high-temperature limit, and are generated from 50 configurations. }
\end{figure}

We also studied the local structure across compositions. In Fig.~\ref{fig:pbte-pbse_bonds}, the distribution of bond lengths is shown across the PbSe$_{1-x}$Te$_{x}$ pseudo-binary. All configurations are weighed equally in these distributions, corresponding to a high-temperature ensemble average.  For PbSe$_{11/12}$Te$_{1/12}$ the bond lengths are tightly clustered around that of PbSe, but the bonds are slightly larger due to the incorporation of the larger Te atom. The amplitude at 3.2 {\AA} originates from Pb-Te bonds, as can be seen in the supplementary. The reverse trend is true for PbSe$_{1/12}$Te$_{11/12}$, where the slight amplitude at 3.15 {\AA} comes from the Pb-Se bonds.  For the compositions towards the middle of the pseudo-binary, the bond distribution has an evidently larger spread. In part, this can be explained by simply having more diversity in the anions. If all Pb-Se and Pb-Te bonds were at a fixed length, then moving to the middle compositions would still increase the spread of the distribution. However, the spread also increases in the Pb-Se and Pb-Te bonds, as can be seen in supplementary Fig.~S1. The position of the largest peak in the distribution shifts linearly as a function of composition. The position of these peaks corresponds to the Vegard's derived bond length. The use of Vegard's law can still show the max peak of the distribution, but more in-depth structural techniques like this method are necessary for ascertaining the extent of structural distortions.

\subsection{{\label{sec:Convex_Hull}PbS$_{x}$Se$_{y}$Te$_{1-x-y}$ pseudo-ternary and the Free energy convex hull}}
\begin{figure*}
\centering
\includegraphics[width=0.8\linewidth]{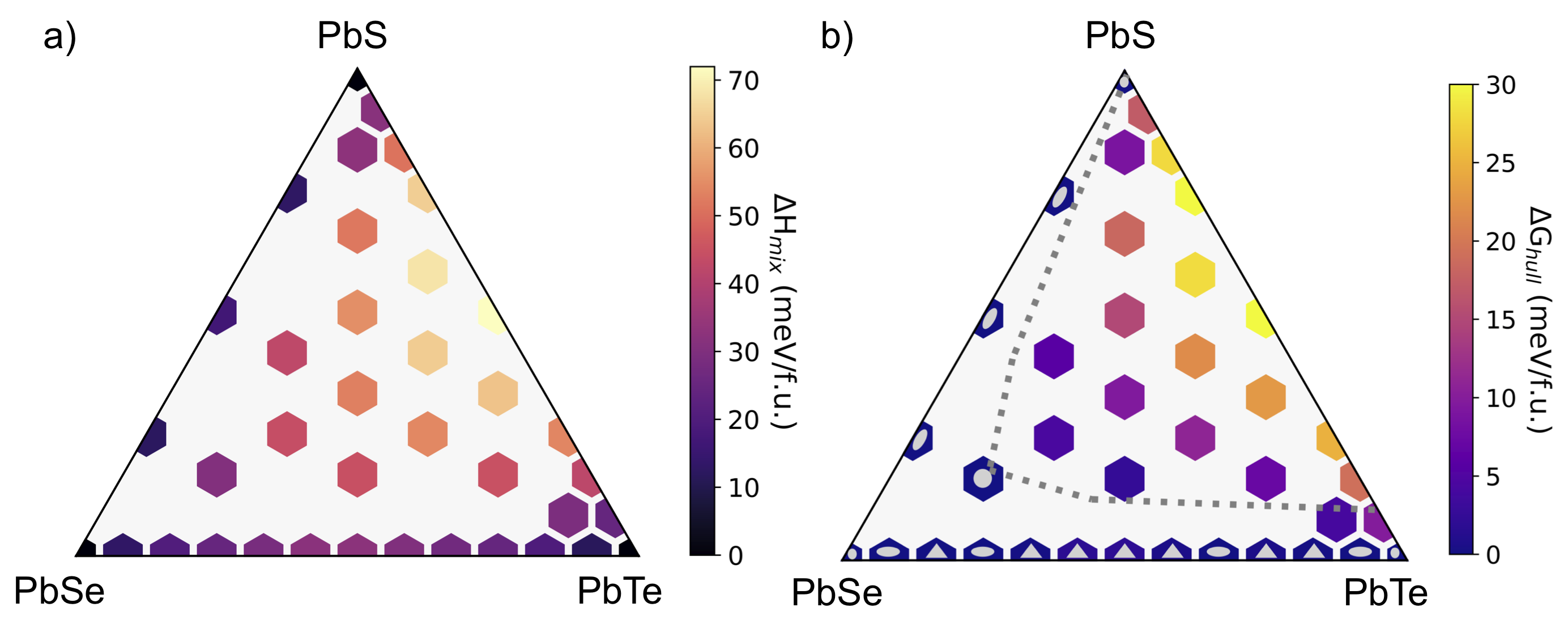}
\caption{\label{fig:ternary}(a) The calculated enthalpy of mixing ($\Delta H_{mix}$) for the PbS$_{x}$Se$_{y}$Te$_{1-x-y}$ space is shown. $\Delta H_{mix}$ is particularly high for PbS-PbTe-rich compositions. (b) The free energy \textit{above} the convex hull ($\Delta G_{hull}$) finds that compositions rich in both PbS and PbTe are at energies above the hull. They are thus subject to decomposition to PbS and PbTe alloys. Grey circles denote being on the hull, and grey triangles signify being that the composition is on the hull within the uncertainty from random sampling (ie. $\Delta G_{hull}<2\sigma_{\Delta H}$). The simulated results are corroborated by experimental findings, although there is a temperature offset. Both $\Delta H_{mix}$ and $\Delta G_{hull}$ were calculated at 600\,K. The grey dashed line is an experimental result showing the region of immiscibility at 773\,K \cite{PbSSeTe-Phase-Diagram}. }

\end{figure*} 

Extending this methodology from psuedobinaries to pseudo-ternaries, we consider the PbS$_{x}$Se$_{y}$Te$_{1-x-y}$ system. 
The enthalpy of mixing ($\Delta H_{mix}$) at 600\,K across this chemical space is shown in Fig.~\ref{fig:ternary}. Within the composition space, PbS$_{0.5}$Te$_{0.5}$ has the highest $\Delta H_{mix}$. This can be explained by the two parent compounds having the largest volume mismatch ($\Delta V=16$ {\AA}$^3$/f.u.), which has been shown to be a significant factor in determining miscibility \cite{hume1935theory,usanmaz2016first}. The $\Delta H_{mix}$ across PbTe-PbSe is lower due to the smaller size mismatch (11 {\AA}$^3$/f.u.), and PbSe-PbS has the lowest $\Delta H_{mix}$, corresponding to its lowest size mismatch (6 {\AA}$^3$/f.u.). 
Adding Se to the PbS-PbTe binary reduces the $\Delta H_{mix}$; one could rationalize this as the Se diluting the unfavorable Te-S interactions. The converse is also true: adding S to the PbSe-PbTe compositions raises its energy by increasing the concentration of S-Te interactions.
\par To determine which compositions of PbS$_{x}$Se$_{y}$Te$_{1-x-y}$ are stable against decomposition, $\Delta G_{mix}$ was calculated for a temperature of 600\,K, and a free energy convex hull was subsequently built from those values. 
The energy difference between the alloy and the free energy convex hull at each point,  $\Delta G_{hull}$, is shown in Fig.~\ref{fig:ternary}b. If a composition is on the hull, that is, the composition is stable against phase separation, then $\Delta G_{hull}=0$. Hexagons with grey dots in them are mathematically on the hull (ie. $\Delta G_{hull}=0$), while hexagons with triangles are on the hull within their uncertainty (ie. $\Delta G_{hull}<2\sigma_{\Delta H}$). Some error is inherent in this approach as the hull is built from discrete points rather than a continuously defined $\Delta G_{mix}$ function.  
We find that there is a wide two-phase region along the PbS-PbTe pseudo-binary. This region narrows as PbSe is added, both by decreasing the enthalpy and increasing the entropy.
These predictions are consistent with prior experimental literature; the dashed line Fig.~\ref{fig:ternary}b reproduces the phase boundary at 773K temperature from reference \cite{PbSSeTe-Phase-Diagram}. The offset of 100-200\,K between experiment and our approach is to be expected. This is likely due to the presented approach ignoring vibrational degrees of freedom. 

\subsection{\label{sec:quinary}(Pb$_{1-x}$Ge$_{x}$)(Te$_{y}$Se$_{z}$S$_{1-y-z}$) High-entropy Space}
\begin{figure*}[t]
\centering
\includegraphics[width=0.9\linewidth]{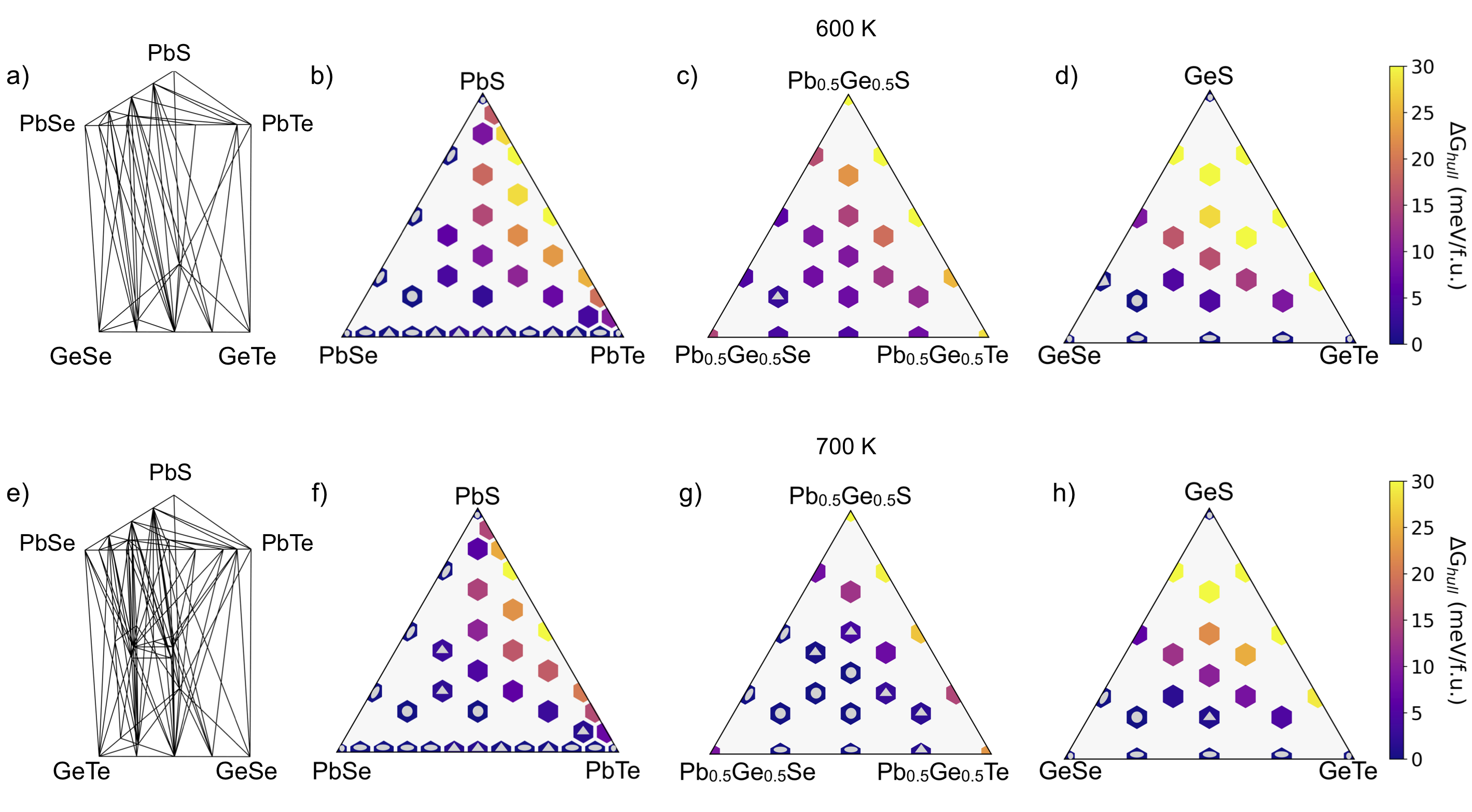}
\caption{(a) The (Pb$_{1-x}$Ge$_{x}$)(Te$_{y}$Se$_{z}$S$_{1-y-z}$) phase space is visualized as a prism with cation composition varying along the axis. 
Pseudo-ternaries (b-d) show three slices of this prism with discrete sampling; the color scheme denotes the free energy above the convex hull. 
Hexagons with a silver point inside indicate that they are on the hull; hexagons with a triangle denote compositions whose distance from the hull is less than their uncertainty (2$\sigma_{\bar{E},n}$, as determined by the CLT).  Calculating the convex hull at 600\,K using the coarse sampling of (b-d) fills the interior of the prism with four-phase tetrahedra. At this temperature, the $x$=0.5 compositions are not on the hull and the resulting tetrahedra of (a) span from $x$=0 to $x$=1 compositions. (e-h) The lower row shows the results at 700\,K. At this temperature, far more compositions are on the hull, including in the interior of the prism (g), where many of the compositions exhibit significant entropy stabilization. }
\label{fig:prism_slices}
\end{figure*} 

We extend the Pb-chalcogenide pseudo-ternary to the high-dimensional (Pb$_{1-x}$Ge$_{x}$)(Te$_{y}$Se$_{z}$S$_{1-y-z}$) system. Here we choose Ge for its small radius relative to Pb, which is expected to result in significant regions of immiscibility. For reference, the ionic radius of Ge is only 39$\%$ smaller than that of Pb when both have a sixfold coordination and a +2 oxidation state \cite{shannon1976revised}. Ge was additionally chosen as an intriguing expansion of the composition space due to the structural diversity found in the end members (i.e.~GeS, GeSe, GeTe). In their ground states, GeS and GeSe are $Pnma$ and GeTe is $R3m$ \cite{GeS,GeSe,GeTe}. Finally, Ge-based chalcogen alloys have shown intriguing thermoelectric behavior \cite{GeTe-Thermoelectrics,GeTeSeS-Thermoelectrics,MGT}. 

In the study of the (Pb$_{1-x}$Ge$_{x}$)(Te$_{y}$Se$_{z}$S$_{1-y-z}$) system, an $n_o$ of 50 configurations with 24-atom supercells was used for all 78 compositions. The standard deviation in the $\Delta H_{mix}$ for a given composition, $\sigma_{H,50}$, remained fairly consistent throughout the composition space. The CLT was employed to calculate the uncertainties in the $\Delta H_{mix}$ using these $\sigma_{H,50}$ values; only Pb$_{0.5}$Ge$_{0.5}$S and GeTe$_{0.25}$S$_{0.75}$ had an uncertainty higher than 3 meV/f.u. Both compositions required only five more configurations to get below the desired uncertainty. 

The (Pb$_{1-x}$Ge$_{x}$)(Te$_{y}$Se$_{z}$S$_{1-y-z}$) composition space yields a pseudo-hexnary alloy represented by a trigonal prism, bounded by two pseudo-ternaries and three pseudo-quaternaries (Fig.~\ref{fig:prism_slices}a).  Concerning the pseudo-ternary faces, the PbS$_{x}$Se$_{y}$Te$_{1-x-y}$ face has been discussed in section \ref{sec:Convex_Hull}.  Fig.~\ref{fig:prism_slices} shows different slices through this composition space at two different temperatures. 

As previously discussed, at $T=600$\,K the pseudo-ternary in Fig.~\ref{fig:prism_slices}b appears to be the most soluble judging by the number of compositions that are on the convex hull. The free energy of the opposite face, GeS$_{x}$Se$_{y}$Te$_{1-x-y}$, is shown in Fig.~\ref{fig:prism_slices} at 600\,K. Once again, the GeS-GeTe alloys are more energetically costly than the GeS-GeSe and GeSe-GeTe alloys. Experimental phase stability measurements indicate extremely limited solubility for GeS-GeTe \cite{GeTeS}. There is also limited solubility for GeS-GeSe \cite{GeSeS}, despite GeS and GeSe having the same crystal structure. Experimental studies of GeSe-GeTe indicate a complete solid solution at temperatures above 930\,K and narrow ranges of immiscibility at low temperature to account for changes in space group \cite{GeTeSe}.  The symmetry-driven regions of immiscibility are fairly narrow, and therefore are not found using the coarse compositional sampling presented here. There is fairly little solubility reported for adding GeSe and GeS to GeTe, with phase separation occurring at the dilute concentration of GeS$_{0.05}$Se$_{0.05}$Te$_{0.9}$ \cite{GeTeSeS-Thermoelectrics}.

The middle slice of the prism (Figure \ref{fig:prism_slices}c) is entirely above the hull at 600\,K. Despite these compositions having the highest entropy in the prism, they are not on the free energy convex hull due to their high $\Delta H_{mix}$ and the presence of competing alloy compositions with low $\Delta G_{mix}$. Highlighting this competition, 9 out of the 22 compositions in the middle slice have negative free energies, and thus would not decompose into their parent compounds, but due to competing alloy compositions, they are not on the hull. For instance, Pb$_{0.5}$Ge${_0.5}$S$_{0.33}$Se$_{0.33}$Te$_{0.34}$ decomposes into GeSe$_{0.25}$Te$_{0.75}$, GeSe$_{0.5}$Te$_{0.5}$, PbSe$_{0.5}$S$_{0.5}$, and PbSe$_{0.25}$S$_{0.75}$. The results would have been qualitatively different if a metric like the temperature of mixing had been used since it only determines the temperature at which $\Delta G_{mix}$ is zero, thus assuming that the parent compounds are the only competing compositions. We find that the presence of competing alloy compositions to be large, especially in high-entropy spaces, underlying the need for a free energy convex hull to determine solubility. The predicted tendency to decompose to GeX- and PbX-rich pseudo-ternaries is consistent with known Pb$_{1-x}$Ge$_{x}$S, Pb$_{1-x}$Ge$_{x}$Se, and Pb$_{1-x}$Ge$_{x}$Te pseudo-binaries \cite{PbGeS,PbGeSe,PbGeTe}. At high temperatures, Pb$_{1-x}$Ge$_{x}$Te is known from experiments to become miscible for a narrow temperature range before melting (843-965\,K) \cite{PbGeTe}.

The interior of the prism shown in Figure \ref{fig:prism_slices}a is completely filled with Alkemade tetrahedra. These four-phase regions arise from determining the convex hull of this three dimensional composition space. It is important to note that because our calculations are limited to single points in composition space, a single phase region can only be represented as a series of adjacent, small, multi-phase regions. This discrete sampling likewise precludes a complete determination of the phase diagram at the current sampling density. However, even this limited sampling provides predictions concerning phase stability. Specifically,  the (Pb$_{0.5}$Ge$_{0.5}$)(Te$_{y}$Se$_{z}$S$_{1-y-z}$) compositions that are above the hull indicate a concave region in the interior of the prism. The edges of this concave region are not, however, well defined as there is likely some mixed cation solubility in the single phase regions at the top and bottom of the prism.
Limited sampling does not provide a lower bound on the energy of the hull, but it does provide an upper bound; if a composition is found to be above the hull, no amount of increased sampling will push it onto the hull. Furthermore, if a composition is found to be on the hull, increased sampling may result in the discovery of a new low energy composition that pushes previously found compositions off the hull.  

Even with the low sampling density considered herein, some preliminary insights into the free energy surface and associated multi-phase regions can be inferred from Figure \ref{fig:prism_slices}a.  The presence of many Alkemade tetrahedra connecting to a single point, as seen along the GeTe-GeSe edge, suggests the presence of a nearby favorable extrema (i.e., large positive Hessian) in the free energy surface. These results are consistent with the experimental results for the pseudo-binaries discussed above. To our knowledge the interior of the prism has not been experimentally studied. 

Considering the 700\,K behavior of Pb$_{1-x}$Ge$_{x}$Te$_{y}$Se$_{z}$S$_{1-y-z}$, we predict significant stabilization of the interior (x = 0.5), as seen in Figure \ref{fig:prism_slices}(e)).  As expected, the lower entropy Ge and Pb pseudo-ternaries show minimal increased stabilization from the 100\,K increase in temperature. The resulting grid is still coarsely sampled (i.e., $x$=0, 0.5, 1) and is likely subject to qualitative changes in the resulting phase diagram with increased density. Nevertheless, the presented method allows for the efficient evaluation of the free energy for many compositions. 

\begin{figure}[t]
\centering
\includegraphics[width = 0.9 \linewidth]{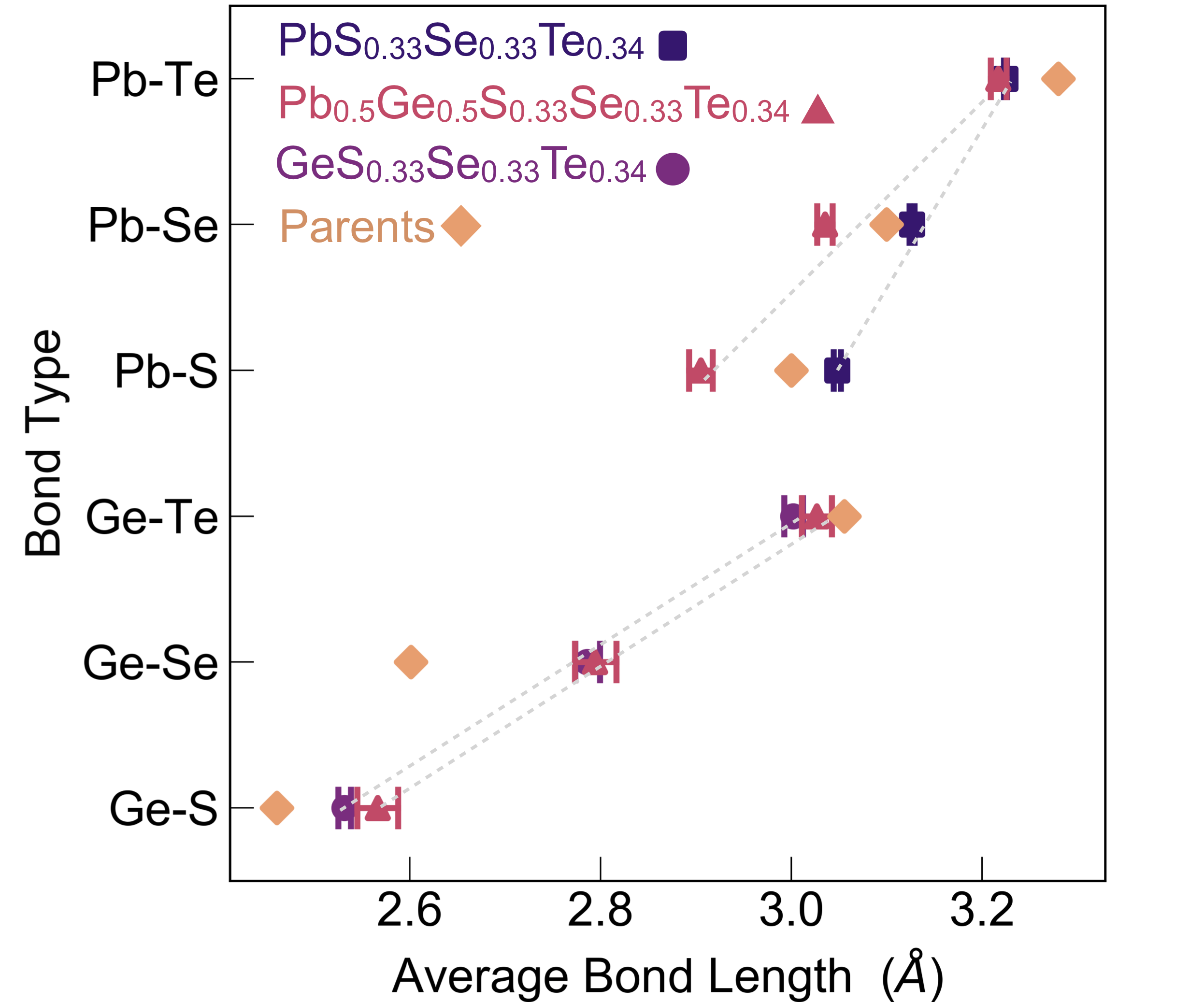}
\caption{\label{fig:HE_Bonds} Local distortion resulting in alloy bond lengths deviating from their parent bond lengths. Further complicatios arise when mixing parent compounds with different structures. Average bond lengths, broken up by type, are shown for various compositions. The bars denote the standard deviation in the bond lengths for a given type and composition. For reference, the parent bond lengths are included as well.}
\end{figure}

To investigate the local structure in the high-entropy alloy, Pb$_{0.5}$Ge$_{0.5}$S$_{0.33}$Se$_{0.33}$Te$_{0.34}$, we 
consider bond lengths  trends (Fig.~\ref{fig:HE_Bonds}) with those of the pseudo-ternaries GeS$_{0.33}$Se$_{0.33}$Te$_{0.34}$ and PbS$_{0.33}$Se$_{0.33}$Te$_{0.34}$ and the parent compounds.
Beginning with PbS$_{0.33}$Se$_{0.33}$Te$_{0.34}$,  the Pb-X bond lengths are close to their parent bond lengths. However, the mixture of anions yields an average lattice constant that drives local strains. 
This strain is visible as the average Pb-Te bond distance shrinks and Pb-S bonds grow such that they are closer to the overall average bond length (3.1 {\AA}). 
As this average bond length is close to that of PbSe,the Pb-Se bonds do not significantly distort.
We also calculated the average coordination number of the composition to be 5.9, indicating the presence of  structural deviations from the undistorted rocksalt with a coordination of 6.

A qualitatively similar trends exist with GeS$_{0.33}$Se$_{0.33}$Te$_{0.34}$; the Ge-Te bonds slightly shrink and the other two bond types have to grow. However, the average coordination number is 4.4, indicating a significant distortion from rocksalt. 
This behavior can be rationalized by examining the parent structure across the Ge-chalcogenides. 
To start, we review  structure and polymorphism in the respective parent structures. Given our algorithm for determining the first coordination shell, GeTe in the $R3m$ space group is found to be six-fold coordinated, and thus the plotted point is the average of its three short bonds (2.85 {\AA}) and three long bonds (3.26 {\AA}). We note that the trend in bond length for the Ge-X compounds is very consistent with the Pb-X compounds if only the short Ge-Te bond is considered in Fig.~\ref{fig:HE_Bonds}. 
This strong distortion can be viewed through the lens of thermodynamics; namely, the $R3m$ structure is 34 meV/f.u. higher in energy than the rocksalt polymorph. GeSe and GeS ($Pnma$) show up as being three-fold coordinated and their rocksalt polymorphs are 20 and 108 meV/f.u. 
Given the above information about the three parents, it is reasonable that even though the Ge atoms start out  octahedrally coordinated in GeS$_{0.33}$Se$_{0.33}$Te$_{0.34}$, they relax locally such that their coordination number is significantly reduced.

With this understanding of the Pb- and Ge-based pseudo-ternary alloys, the behavior of Pb$_{0.5}$Ge${_0.5}$S$_{0.33}$Se$_{0.33}$Te$_{0.34}$ can be rationalized. The average coordination number is 4.3 within this alloy and is these broken bonds are evenly distributed between Ge and Pb; this is quite low given the nearly perfect octahedral coordination found in PbS$_{0.33}$Se$_{0.33}$Te$_{0.34}$.
The average bond length in this compound is 3.0 {\AA}; this leads to a shrinking of the Pb-S and Pb-Se bond lengths.  In contrast, the  distribution of Pb-Te bond lengths are relatively unaffected by the addition of Ge at high concentrations.  
Considering the Ge-X bonds within Pb$_{0.5}$Ge$_{0.5}$S$_{0.33}$Se$_{0.33}$Te$_{0.34}$, the addition of Pb is largely inconsequential. The average coordination number is 4.2 and and the bond lengths are consistent with the GeS$_{0.33}$Se$_{0.33}$Te$_{0.34}$ pseudoternary.  

It is intriguing that despite the large structural distortions away from the Pb-chalcogenide parents, Pb$_{0.5}$Ge$_{0.5}$S$_{0.33}$Se$_{0.33}$Te$_{0.34}$ is on the free energy convex hull at 700\,K. The result is a testament to the magnitude of stabilization that is possible from entropy, and the interesting structures that entropy stabilization enable.  

\section{Discussion}
Having demonstrated the utility of this method in a variety of alloy spaces, here we review the strengths, identify persistent challenges, and highlight opportunities for expanding upon the current work. Broadly, we try to position this method in between two extremes for simulating alloys--using a single large supercell like SQS, and completely sampling small supercells, as has been done with other implentations of the independent cell approximation \cite{aflowpocc,aflowefa}. Working within the independent cell approximation allows for the evaluation of ensemble properties as a function of synthesis temperature. By employing random sampling, we are able to evaluate larger, more disordered supercells in a computationally efficient manner. Using these supercells allows for an in-depth evaluation of the local structure, all as a function of synthesis temperature. Lastly, random sampling provides the basis for effectively navigating the computational trade-offs between precision and computational cost; from the Central Limit Theorem, we derive the relationship between precision and the number of additional samples. Doing so allows for a judicious use of computational resources in exploring composition space. Furthermore, having the ability to trade off between cost and uncertainty can be leveraged in optimal experimental design \cite{settles2009active} and Bayesian optimization methods \cite{garnett2022bayesian}.

Nevertheless, effective use of this method requires care. It is important to be aware of polymorphism in alloys and rare configurational ground states that might not be found using limited random sampling. Furthermore, any thermodynamic methodology will suffer from the fact that there could be competing stoichiometries that push compositions off the hull. 

\textbf{Polymorphism.} In the efforts herein, the polymorphic competition for the parent binaries is between three phases (rocksalt, $R3m$--a rhomboheral perturbation, and the more significant $Pnma$ distortion). Within DFT relaxations, we have found continuous transitions between these structure-types and the ergodic hypothesis is thus generally satisfied. 
However, caution must be taken in initializing heterostructural alloys where the parents have different ground state structures without continuous transitions between them (e.g., zincblende and rocksalt). Here, relaxations in DFT may only capture the local ground state structure-type and miss the global ground state. Testing the sensitivity to starting lattice choice is, in these cases, critical. Further, \textit{post hoc} grouping of the resulting structures based on space group and local coordination is needed to avoid incorrectly assuming ergodicity between all sampled structure-types \cite{jones2017polymorphism,stevanovic2016sampling}. Another method for comparing free energies of various structure-types is to start with amorphous structures rather than known crystalline prototypes, and conduct structural relaxations, as we have done in various other works \cite{jones2017polymorphism,stevanovic2016sampling}. This approach provides insight into the basins of attraction for these polymorphs and makes no assumptions about structure-type. 

\textbf{Sampling Limitations.} In importance sampling, the target probability distribution is not directly accessible; in our case, this would be the probability of any specific configuration occurring. Instead, we randomly sample from a uniform distribution of configurations; in the event that the true probability distribution is close to random, then our method will converge with relatively less samples than a highly non-uniform distribution \cite{importance}. In the high-temperature limit, the true probability distribution is uniform, and at the temperatures of interest for our systems, the probability distribution is nearly uniform. As such, random sampling is an effective method to assemble an ensemble. For systems equilibrated at low temperature, or have strong ordering tendencies, a larger number of samples will be required to accurately represent the distribution. Importantly, there are two sources of error. The first is due to incompletely sampling from a distribution of configurations, and the second is due to sampling from an erroneous distribution. The Central Limit Theorem allows for the evaluation of the first source of error, but not the second. Thus, the Central Limit Theorem may offer misleading guidance with respect to convergence in this situation.  

The challenge of highly non-uniform distributions can be addressed in a multitude of ways. First, complete sampling using small supercells can be used to partially mitigate this risk. Second, Monte Carlo (MC) can be used to sample from an approximate distribution that more closely resembles the true distribution \cite{wallace2019atomistic,cordell2021probing,wolverton1998first,van2002automating}. MC inherently requires discarding a portion of the overall calculations, and thus requires a model Hamiltonian to make the computational cost accessible. This second point underlines how the independent supercell approach is not antithetical to model Hamiltonian-based techniques; indeed, pairing these approaches would allow for efficient sampling from highly non-uniform distributions.  

\section{Conclusion}
Evaluating the stability and structure across high dimensional alloy space is fundamental to future high throughput searches for high performing alloys.
In this work, we implemented the independent supercell approximation, allowing for the calculation of the thermodynamics and structure of disordered alloys. 
Applying this method to the (Pb,Ge)(S,Se,Te) composition space and sub-spaces therein, we predicted phase diagrams consistent with prior experimental literature, and make novel predictions concerning the local structure. 
Consideration of the convex hull reveals that some high entropy compositions (e.g.~Pb$_{0.5}$Ge$_{0.5}$S$_{0.33}$Se$_{0.33}$Te$_{0.34}$, Pb$_{0.5}$Ge$_{0.5}$S$_{0.16}$Se$_{0.67}$Te$_{0.17}$) are stable at high temperatures despite their highly distorted structure and large $\Delta H_{mix}$.  
The presented method can also be extended to ensemble properties, as initially demonstrated by the consideration of the PbSe$_{0.5}$Te$_{0.5}$ bulk modulus. 

\section{Acknowledgements}
This material is based upon work supported by the National Science Foundation under Grant No.~1729594. V.S. acknowledges support of the National Science Foundation under Grant No.~1945010. E.T. acknowledges the support of the National Science Foundation under Grant No.~1940199 and 2118201. R.G. also recognizes the support of the National Science Foundation under Grant No.~1940224. The use of the Colorado School of Mines's computing resources is gratefully acknowledged.

\bibliography{bib}

\providecommand{\noopsort}[1]{}\providecommand{\singleletter}[1]{#1}%
\begin{thebibliography}{67}%
\makeatletter
\providecommand \@ifxundefined [1]{%
 \@ifx{#1\undefined}
}%
\providecommand \@ifnum [1]{%
 \ifnum #1\expandafter \@firstoftwo
 \else \expandafter \@secondoftwo
 \fi
}%
\providecommand \@ifx [1]{%
 \ifx #1\expandafter \@firstoftwo
 \else \expandafter \@secondoftwo
 \fi
}%
\providecommand \natexlab [1]{#1}%
\providecommand \enquote  [1]{``#1''}%
\providecommand \bibnamefont  [1]{#1}%
\providecommand \bibfnamefont [1]{#1}%
\providecommand \citenamefont [1]{#1}%
\providecommand \href@noop [0]{\@secondoftwo}%
\providecommand \href [0]{\begingroup \@sanitize@url \@href}%
\providecommand \@href[1]{\@@startlink{#1}\@@href}%
\providecommand \@@href[1]{\endgroup#1\@@endlink}%
\providecommand \@sanitize@url [0]{\catcode `\\12\catcode `\$12\catcode
  `\&12\catcode `\#12\catcode `\^12\catcode `\_12\catcode `\%12\relax}%
\providecommand \@@startlink[1]{}%
\providecommand \@@endlink[0]{}%
\providecommand \url  [0]{\begingroup\@sanitize@url \@url }%
\providecommand \@url [1]{\endgroup\@href {#1}{\urlprefix }}%
\providecommand \urlprefix  [0]{URL }%
\providecommand \Eprint [0]{\href }%
\providecommand \doibase [0]{https://doi.org/}%
\providecommand \selectlanguage [0]{\@gobble}%
\providecommand \bibinfo  [0]{\@secondoftwo}%
\providecommand \bibfield  [0]{\@secondoftwo}%
\providecommand \translation [1]{[#1]}%
\providecommand \BibitemOpen [0]{}%
\providecommand \bibitemStop [0]{}%
\providecommand \bibitemNoStop [0]{.\EOS\space}%
\providecommand \EOS [0]{\spacefactor3000\relax}%
\providecommand \BibitemShut  [1]{\csname bibitem#1\endcsname}%
\let\auto@bib@innerbib\@empty
\bibitem [{\citenamefont {Schnepf}\ \emph {et~al.}(2020)\citenamefont
  {Schnepf}, \citenamefont {Cordell}, \citenamefont {Tellekamp}, \citenamefont
  {Melamed}, \citenamefont {Greenaway}, \citenamefont {Mis}, \citenamefont
  {Brennecka}, \citenamefont {Christensen}, \citenamefont {Tucker},
  \citenamefont {Toberer}, \citenamefont {Lany},\ and\ \citenamefont
  {Tamboli}}]{schnepf2020utilizing}%
  \BibitemOpen
  \bibfield  {author} {\bibinfo {author} {\bibfnamefont {R.~R.}\ \bibnamefont
  {Schnepf}}, \bibinfo {author} {\bibfnamefont {J.~J.}\ \bibnamefont
  {Cordell}}, \bibinfo {author} {\bibfnamefont {M.~B.}\ \bibnamefont
  {Tellekamp}}, \bibinfo {author} {\bibfnamefont {C.~L.}\ \bibnamefont
  {Melamed}}, \bibinfo {author} {\bibfnamefont {A.~L.}\ \bibnamefont
  {Greenaway}}, \bibinfo {author} {\bibfnamefont {A.}~\bibnamefont {Mis}},
  \bibinfo {author} {\bibfnamefont {G.~L.}\ \bibnamefont {Brennecka}}, \bibinfo
  {author} {\bibfnamefont {S.}~\bibnamefont {Christensen}}, \bibinfo {author}
  {\bibfnamefont {G.~J.}\ \bibnamefont {Tucker}}, \bibinfo {author}
  {\bibfnamefont {E.~S.}\ \bibnamefont {Toberer}}, \bibinfo {author}
  {\bibfnamefont {S.}~\bibnamefont {Lany}},\ and\ \bibinfo {author}
  {\bibfnamefont {A.~C.}\ \bibnamefont {Tamboli}},\ }\bibfield  {title}
  {\bibinfo {title} {Utilizing site disorder in the development of new
  energy-relevant semiconductors},\ }\href@noop {} {\bibfield  {journal}
  {\bibinfo  {journal} {ACS Energy Letters}\ }\textbf {\bibinfo {volume} {5}},\
  \bibinfo {pages} {2027} (\bibinfo {year} {2020})}\BibitemShut {NoStop}%
\bibitem [{\citenamefont {King}\ \emph {et~al.}(2007)\citenamefont {King},
  \citenamefont {Law}, \citenamefont {Edmondson}, \citenamefont {Fetzer},
  \citenamefont {Kinsey}, \citenamefont {Yoon}, \citenamefont {Sherif},\ and\
  \citenamefont {Karam}}]{king2007solaralloy}%
  \BibitemOpen
  \bibfield  {author} {\bibinfo {author} {\bibfnamefont {R.}~\bibnamefont
  {King}}, \bibinfo {author} {\bibfnamefont {a.}~\bibnamefont {Law}}, \bibinfo
  {author} {\bibfnamefont {K.}~\bibnamefont {Edmondson}}, \bibinfo {author}
  {\bibfnamefont {C.}~\bibnamefont {Fetzer}}, \bibinfo {author} {\bibfnamefont
  {G.}~\bibnamefont {Kinsey}}, \bibinfo {author} {\bibfnamefont
  {H.}~\bibnamefont {Yoon}}, \bibinfo {author} {\bibfnamefont {R.}~\bibnamefont
  {Sherif}},\ and\ \bibinfo {author} {\bibfnamefont {N.}~\bibnamefont
  {Karam}},\ }\bibfield  {title} {\bibinfo {title} {40\% efficient metamorphic
  gainp/ gainas/ ge multijunction solar cells},\ }\href@noop {} {\bibfield
  {journal} {\bibinfo  {journal} {Applied physics letters}\ }\textbf {\bibinfo
  {volume} {90}},\ \bibinfo {pages} {183516} (\bibinfo {year}
  {2007})}\BibitemShut {NoStop}%
\bibitem [{\citenamefont {Geisz}\ \emph {et~al.}(2020)\citenamefont {Geisz},
  \citenamefont {France}, \citenamefont {Schulte}, \citenamefont {Steiner},
  \citenamefont {Norman}, \citenamefont {Guthrey}, \citenamefont {Young},
  \citenamefont {Song},\ and\ \citenamefont {Moriarty}}]{sixjunction2020}%
  \BibitemOpen
  \bibfield  {author} {\bibinfo {author} {\bibfnamefont {J.~F.}\ \bibnamefont
  {Geisz}}, \bibinfo {author} {\bibfnamefont {R.~M.}\ \bibnamefont {France}},
  \bibinfo {author} {\bibfnamefont {K.~L.}\ \bibnamefont {Schulte}}, \bibinfo
  {author} {\bibfnamefont {M.~A.}\ \bibnamefont {Steiner}}, \bibinfo {author}
  {\bibfnamefont {A.~G.}\ \bibnamefont {Norman}}, \bibinfo {author}
  {\bibfnamefont {H.~L.}\ \bibnamefont {Guthrey}}, \bibinfo {author}
  {\bibfnamefont {M.~R.}\ \bibnamefont {Young}}, \bibinfo {author}
  {\bibfnamefont {T.}~\bibnamefont {Song}},\ and\ \bibinfo {author}
  {\bibfnamefont {T.}~\bibnamefont {Moriarty}},\ }\bibfield  {title} {\bibinfo
  {title} {Six-junction iii--v solar cells with 47.1\% conversion efficiency
  under 143 suns concentration},\ }\href@noop {} {\bibfield  {journal}
  {\bibinfo  {journal} {Nature energy}\ }\textbf {\bibinfo {volume} {5}},\
  \bibinfo {pages} {326} (\bibinfo {year} {2020})}\BibitemShut {NoStop}%
\bibitem [{\citenamefont {Witting}\ \emph {et~al.}(2019)\citenamefont
  {Witting}, \citenamefont {Chasapis}, \citenamefont {Ricci}, \citenamefont
  {Peters}, \citenamefont {Heinz}, \citenamefont {Hautier},\ and\ \citenamefont
  {Snyder}}]{Witting2019HEalloy}%
  \BibitemOpen
  \bibfield  {author} {\bibinfo {author} {\bibfnamefont {I.~T.}\ \bibnamefont
  {Witting}}, \bibinfo {author} {\bibfnamefont {T.~C.}\ \bibnamefont
  {Chasapis}}, \bibinfo {author} {\bibfnamefont {F.}~\bibnamefont {Ricci}},
  \bibinfo {author} {\bibfnamefont {M.}~\bibnamefont {Peters}}, \bibinfo
  {author} {\bibfnamefont {N.~A.}\ \bibnamefont {Heinz}}, \bibinfo {author}
  {\bibfnamefont {G.}~\bibnamefont {Hautier}},\ and\ \bibinfo {author}
  {\bibfnamefont {G.~J.}\ \bibnamefont {Snyder}},\ }\bibfield  {title}
  {\bibinfo {title} {The thermoelectric properties of bismuth telluride},\
  }\href@noop {} {\bibfield  {journal} {\bibinfo  {journal} {Adv. Electron.
  Mater.}\ }\textbf {\bibinfo {volume} {5}},\ \bibinfo {pages} {1800904}
  (\bibinfo {year} {2019})}\BibitemShut {NoStop}%
\bibitem [{\citenamefont {Oses}\ \emph {et~al.}(2020)\citenamefont {Oses},
  \citenamefont {Toher},\ and\ \citenamefont {Curtarolo}}]{oses2020high}%
  \BibitemOpen
  \bibfield  {author} {\bibinfo {author} {\bibfnamefont {C.}~\bibnamefont
  {Oses}}, \bibinfo {author} {\bibfnamefont {C.}~\bibnamefont {Toher}},\ and\
  \bibinfo {author} {\bibfnamefont {S.}~\bibnamefont {Curtarolo}},\ }\bibfield
  {title} {\bibinfo {title} {High-entropy ceramics},\ }\href@noop {} {\bibfield
   {journal} {\bibinfo  {journal} {Nature Reviews Materials}\ }\textbf
  {\bibinfo {volume} {5}},\ \bibinfo {pages} {295} (\bibinfo {year}
  {2020})}\BibitemShut {NoStop}%
\bibitem [{\citenamefont {Zunger}\ \emph {et~al.}(1990)\citenamefont {Zunger},
  \citenamefont {Wei}, \citenamefont {Ferreira},\ and\ \citenamefont
  {Bernard}}]{zunger1990special}%
  \BibitemOpen
  \bibfield  {author} {\bibinfo {author} {\bibfnamefont {A.}~\bibnamefont
  {Zunger}}, \bibinfo {author} {\bibfnamefont {S.-H.}\ \bibnamefont {Wei}},
  \bibinfo {author} {\bibfnamefont {L.}~\bibnamefont {Ferreira}},\ and\
  \bibinfo {author} {\bibfnamefont {J.~E.}\ \bibnamefont {Bernard}},\
  }\bibfield  {title} {\bibinfo {title} {Special quasirandom structures},\
  }\href@noop {} {\bibfield  {journal} {\bibinfo  {journal} {Physical review
  letters}\ }\textbf {\bibinfo {volume} {65}},\ \bibinfo {pages} {353}
  (\bibinfo {year} {1990})}\BibitemShut {NoStop}%
\bibitem [{\citenamefont {Ferreira}\ \emph {et~al.}(1989)\citenamefont
  {Ferreira}, \citenamefont {Wei},\ and\ \citenamefont
  {Zunger}}]{ferreira1989first}%
  \BibitemOpen
  \bibfield  {author} {\bibinfo {author} {\bibfnamefont {L.}~\bibnamefont
  {Ferreira}}, \bibinfo {author} {\bibfnamefont {S.-H.}\ \bibnamefont {Wei}},\
  and\ \bibinfo {author} {\bibfnamefont {A.}~\bibnamefont {Zunger}},\
  }\bibfield  {title} {\bibinfo {title} {First-principles calculation of alloy
  phase diagrams: The renormalized-interaction approach},\ }\href@noop {}
  {\bibfield  {journal} {\bibinfo  {journal} {Physical Review B}\ }\textbf
  {\bibinfo {volume} {40}},\ \bibinfo {pages} {3197} (\bibinfo {year}
  {1989})}\BibitemShut {NoStop}%
\bibitem [{\citenamefont {Jiang}\ and\ \citenamefont
  {Uberuaga}(2016)}]{independent-supercell-alloy-2016}%
  \BibitemOpen
  \bibfield  {author} {\bibinfo {author} {\bibfnamefont {C.}~\bibnamefont
  {Jiang}}\ and\ \bibinfo {author} {\bibfnamefont {B.~P.}\ \bibnamefont
  {Uberuaga}},\ }\bibfield  {title} {\bibinfo {title} {Efficient ab initio
  modeling of random multicomponent alloys},\ }\href@noop {} {\bibfield
  {journal} {\bibinfo  {journal} {Physical review letters}\ }\textbf {\bibinfo
  {volume} {116}},\ \bibinfo {pages} {105501} (\bibinfo {year}
  {2016})}\BibitemShut {NoStop}%
\bibitem [{\citenamefont {Pomrehn}\ \emph {et~al.}(2011)\citenamefont
  {Pomrehn}, \citenamefont {Toberer}, \citenamefont {Snyder},\ and\
  \citenamefont {Van De~Walle}}]{pomrehn2011entropic}%
  \BibitemOpen
  \bibfield  {author} {\bibinfo {author} {\bibfnamefont {G.~S.}\ \bibnamefont
  {Pomrehn}}, \bibinfo {author} {\bibfnamefont {E.~S.}\ \bibnamefont
  {Toberer}}, \bibinfo {author} {\bibfnamefont {G.~J.}\ \bibnamefont
  {Snyder}},\ and\ \bibinfo {author} {\bibfnamefont {A.}~\bibnamefont {Van
  De~Walle}},\ }\bibfield  {title} {\bibinfo {title} {Entropic stabilization
  and retrograde solubility in zn 4 sb 3},\ }\href@noop {} {\bibfield
  {journal} {\bibinfo  {journal} {Physical Review B}\ }\textbf {\bibinfo
  {volume} {83}},\ \bibinfo {pages} {094106} (\bibinfo {year}
  {2011})}\BibitemShut {NoStop}%
\bibitem [{\citenamefont {Biswas}\ and\ \citenamefont
  {Lany}(2009)}]{lany-anti-SQS}%
  \BibitemOpen
  \bibfield  {author} {\bibinfo {author} {\bibfnamefont {K.}~\bibnamefont
  {Biswas}}\ and\ \bibinfo {author} {\bibfnamefont {S.}~\bibnamefont {Lany}},\
  }\bibfield  {title} {\bibinfo {title} {Energetics of quaternary iii-v alloys
  described by incorporation and clustering of impurities},\ }\href@noop {}
  {\bibfield  {journal} {\bibinfo  {journal} {Physical Review B}\ }\textbf
  {\bibinfo {volume} {80}},\ \bibinfo {pages} {115206} (\bibinfo {year}
  {2009})}\BibitemShut {NoStop}%
\bibitem [{\citenamefont {Keen}\ and\ \citenamefont
  {Goodwin}(2015)}]{keen2015crystallography}%
  \BibitemOpen
  \bibfield  {author} {\bibinfo {author} {\bibfnamefont {D.~A.}\ \bibnamefont
  {Keen}}\ and\ \bibinfo {author} {\bibfnamefont {A.~L.}\ \bibnamefont
  {Goodwin}},\ }\bibfield  {title} {\bibinfo {title} {The crystallography of
  correlated disorder},\ }\href@noop {} {\bibfield  {journal} {\bibinfo
  {journal} {Nature}\ }\textbf {\bibinfo {volume} {521}},\ \bibinfo {pages}
  {303} (\bibinfo {year} {2015})}\BibitemShut {NoStop}%
\bibitem [{\citenamefont {Simonov}\ and\ \citenamefont
  {Goodwin}(2020)}]{simonov2020designing}%
  \BibitemOpen
  \bibfield  {author} {\bibinfo {author} {\bibfnamefont {A.}~\bibnamefont
  {Simonov}}\ and\ \bibinfo {author} {\bibfnamefont {A.~L.}\ \bibnamefont
  {Goodwin}},\ }\bibfield  {title} {\bibinfo {title} {Designing disorder into
  crystalline materials},\ }\href@noop {} {\bibfield  {journal} {\bibinfo
  {journal} {Nature Reviews Chemistry}\ }\textbf {\bibinfo {volume} {4}},\
  \bibinfo {pages} {657} (\bibinfo {year} {2020})}\BibitemShut {NoStop}%
\bibitem [{\citenamefont {Cordell}\ \emph {et~al.}(2021)\citenamefont
  {Cordell}, \citenamefont {Pan}, \citenamefont {Tamboli}, \citenamefont
  {Tucker},\ and\ \citenamefont {Lany}}]{cordell2021probing}%
  \BibitemOpen
  \bibfield  {author} {\bibinfo {author} {\bibfnamefont {J.~J.}\ \bibnamefont
  {Cordell}}, \bibinfo {author} {\bibfnamefont {J.}~\bibnamefont {Pan}},
  \bibinfo {author} {\bibfnamefont {A.~C.}\ \bibnamefont {Tamboli}}, \bibinfo
  {author} {\bibfnamefont {G.~J.}\ \bibnamefont {Tucker}},\ and\ \bibinfo
  {author} {\bibfnamefont {S.}~\bibnamefont {Lany}},\ }\bibfield  {title}
  {\bibinfo {title} {Probing configurational disorder in zngen 2 using
  cluster-based monte carlo},\ }\href@noop {} {\bibfield  {journal} {\bibinfo
  {journal} {Physical Review Materials}\ }\textbf {\bibinfo {volume} {5}},\
  \bibinfo {pages} {024604} (\bibinfo {year} {2021})}\BibitemShut {NoStop}%
\bibitem [{\citenamefont {Wolverton}\ \emph {et~al.}(1998)\citenamefont
  {Wolverton}, \citenamefont {Ozoli{\c{n}}{\v{s}}},\ and\ \citenamefont
  {Zunger}}]{wolverton1998first}%
  \BibitemOpen
  \bibfield  {author} {\bibinfo {author} {\bibfnamefont {C.}~\bibnamefont
  {Wolverton}}, \bibinfo {author} {\bibfnamefont {V.}~\bibnamefont
  {Ozoli{\c{n}}{\v{s}}}},\ and\ \bibinfo {author} {\bibfnamefont
  {A.}~\bibnamefont {Zunger}},\ }\bibfield  {title} {\bibinfo {title}
  {First-principles theory of short-range order in size-mismatched metal
  alloys: Cu-au, cu-ag, and ni-au},\ }\href@noop {} {\bibfield  {journal}
  {\bibinfo  {journal} {Physical Review B}\ }\textbf {\bibinfo {volume} {57}},\
  \bibinfo {pages} {4332} (\bibinfo {year} {1998})}\BibitemShut {NoStop}%
\bibitem [{\citenamefont {van~de Walle}\ and\ \citenamefont
  {Ceder}(2002)}]{van2002automating}%
  \BibitemOpen
  \bibfield  {author} {\bibinfo {author} {\bibfnamefont {A.}~\bibnamefont
  {van~de Walle}}\ and\ \bibinfo {author} {\bibfnamefont {G.}~\bibnamefont
  {Ceder}},\ }\bibfield  {title} {\bibinfo {title} {Automating first-principles
  phase diagram calculations},\ }\href@noop {} {\bibfield  {journal} {\bibinfo
  {journal} {Journal of Phase Equilibria}\ }\textbf {\bibinfo {volume} {23}},\
  \bibinfo {pages} {348} (\bibinfo {year} {2002})}\BibitemShut {NoStop}%
\bibitem [{\citenamefont {Zunger}\ and\ \citenamefont
  {Jaffe}(1983)}]{zunger1983structural}%
  \BibitemOpen
  \bibfield  {author} {\bibinfo {author} {\bibfnamefont {A.}~\bibnamefont
  {Zunger}}\ and\ \bibinfo {author} {\bibfnamefont {J.}~\bibnamefont {Jaffe}},\
  }\bibfield  {title} {\bibinfo {title} {Structural origin of optical bowing in
  semiconductor alloys},\ }\href@noop {} {\bibfield  {journal} {\bibinfo
  {journal} {Physical Review Letters}\ }\textbf {\bibinfo {volume} {51}},\
  \bibinfo {pages} {662} (\bibinfo {year} {1983})}\BibitemShut {NoStop}%
\bibitem [{\citenamefont {Ortiz}\ \emph {et~al.}(2015)\citenamefont {Ortiz},
  \citenamefont {Peng}, \citenamefont {Lopez}, \citenamefont {Parilla},
  \citenamefont {Lany},\ and\ \citenamefont {Toberer}}]{ortiz2015effect}%
  \BibitemOpen
  \bibfield  {author} {\bibinfo {author} {\bibfnamefont {B.~R.}\ \bibnamefont
  {Ortiz}}, \bibinfo {author} {\bibfnamefont {H.}~\bibnamefont {Peng}},
  \bibinfo {author} {\bibfnamefont {A.}~\bibnamefont {Lopez}}, \bibinfo
  {author} {\bibfnamefont {P.~A.}\ \bibnamefont {Parilla}}, \bibinfo {author}
  {\bibfnamefont {S.}~\bibnamefont {Lany}},\ and\ \bibinfo {author}
  {\bibfnamefont {E.~S.}\ \bibnamefont {Toberer}},\ }\bibfield  {title}
  {\bibinfo {title} {Effect of extended strain fields on point defect phonon
  scattering in thermoelectric materials},\ }\href@noop {} {\bibfield
  {journal} {\bibinfo  {journal} {Physical Chemistry Chemical Physics}\
  }\textbf {\bibinfo {volume} {17}},\ \bibinfo {pages} {19410} (\bibinfo {year}
  {2015})}\BibitemShut {NoStop}%
\bibitem [{\citenamefont {Gurunathan}\ \emph {et~al.}(2020)\citenamefont
  {Gurunathan}, \citenamefont {Hanus},\ and\ \citenamefont
  {Snyder}}]{gurunathan2020alloy}%
  \BibitemOpen
  \bibfield  {author} {\bibinfo {author} {\bibfnamefont {R.}~\bibnamefont
  {Gurunathan}}, \bibinfo {author} {\bibfnamefont {R.}~\bibnamefont {Hanus}},\
  and\ \bibinfo {author} {\bibfnamefont {G.~J.}\ \bibnamefont {Snyder}},\
  }\bibfield  {title} {\bibinfo {title} {Alloy scattering of phonons},\
  }\href@noop {} {\bibfield  {journal} {\bibinfo  {journal} {Materials
  Horizons}\ }\textbf {\bibinfo {volume} {7}},\ \bibinfo {pages} {1452}
  (\bibinfo {year} {2020})}\BibitemShut {NoStop}%
\bibitem [{\citenamefont {Gurunathan}\ \emph {et~al.}(2022)\citenamefont
  {Gurunathan}, \citenamefont {Sarker}, \citenamefont {Borg}, \citenamefont
  {Saal}, \citenamefont {Ward}, \citenamefont {Mehta},\ and\ \citenamefont
  {Snyder}}]{gurunathan2022mapping}%
  \BibitemOpen
  \bibfield  {author} {\bibinfo {author} {\bibfnamefont {R.}~\bibnamefont
  {Gurunathan}}, \bibinfo {author} {\bibfnamefont {S.}~\bibnamefont {Sarker}},
  \bibinfo {author} {\bibfnamefont {C.~K.}\ \bibnamefont {Borg}}, \bibinfo
  {author} {\bibfnamefont {J.}~\bibnamefont {Saal}}, \bibinfo {author}
  {\bibfnamefont {L.}~\bibnamefont {Ward}}, \bibinfo {author} {\bibfnamefont
  {A.}~\bibnamefont {Mehta}},\ and\ \bibinfo {author} {\bibfnamefont {G.~J.}\
  \bibnamefont {Snyder}},\ }\bibfield  {title} {\bibinfo {title} {Mapping
  thermoelectric transport in a multicomponent alloy space},\ }\href@noop {}
  {\bibfield  {journal} {\bibinfo  {journal} {arXiv preprint arXiv:2205.01520}\
  } (\bibinfo {year} {2022})}\BibitemShut {NoStop}%
\bibitem [{\citenamefont {Yang}\ \emph {et~al.}(2016)\citenamefont {Yang},
  \citenamefont {Oses},\ and\ \citenamefont {Curtarolo}}]{aflowpocc}%
  \BibitemOpen
  \bibfield  {author} {\bibinfo {author} {\bibfnamefont {K.}~\bibnamefont
  {Yang}}, \bibinfo {author} {\bibfnamefont {C.}~\bibnamefont {Oses}},\ and\
  \bibinfo {author} {\bibfnamefont {S.}~\bibnamefont {Curtarolo}},\ }\bibfield
  {title} {\bibinfo {title} {Modeling off-stoichiometry materials with a
  high-throughput ab-initio approach},\ }\href@noop {} {\bibfield  {journal}
  {\bibinfo  {journal} {Chemistry of Materials}\ }\textbf {\bibinfo {volume}
  {28}},\ \bibinfo {pages} {6484} (\bibinfo {year} {2016})}\BibitemShut
  {NoStop}%
\bibitem [{\citenamefont {Lederer}\ \emph {et~al.}(2018)\citenamefont
  {Lederer}, \citenamefont {Toher}, \citenamefont {Vecchio},\ and\
  \citenamefont {Curtarolo}}]{aflowlvtc}%
  \BibitemOpen
  \bibfield  {author} {\bibinfo {author} {\bibfnamefont {Y.}~\bibnamefont
  {Lederer}}, \bibinfo {author} {\bibfnamefont {C.}~\bibnamefont {Toher}},
  \bibinfo {author} {\bibfnamefont {K.~S.}\ \bibnamefont {Vecchio}},\ and\
  \bibinfo {author} {\bibfnamefont {S.}~\bibnamefont {Curtarolo}},\ }\bibfield
  {title} {\bibinfo {title} {The search for high entropy alloys: a
  high-throughput ab-initio approach},\ }\href@noop {} {\bibfield  {journal}
  {\bibinfo  {journal} {Acta Materialia}\ }\textbf {\bibinfo {volume} {159}},\
  \bibinfo {pages} {364} (\bibinfo {year} {2018})}\BibitemShut {NoStop}%
\bibitem [{\citenamefont {Sarker}\ \emph {et~al.}(2018)\citenamefont {Sarker},
  \citenamefont {Harrington}, \citenamefont {Toher}, \citenamefont {Oses},
  \citenamefont {Samiee}, \citenamefont {Maria}, \citenamefont {Brenner},
  \citenamefont {Vecchio},\ and\ \citenamefont {Curtarolo}}]{aflowefa}%
  \BibitemOpen
  \bibfield  {author} {\bibinfo {author} {\bibfnamefont {P.}~\bibnamefont
  {Sarker}}, \bibinfo {author} {\bibfnamefont {T.}~\bibnamefont {Harrington}},
  \bibinfo {author} {\bibfnamefont {C.}~\bibnamefont {Toher}}, \bibinfo
  {author} {\bibfnamefont {C.}~\bibnamefont {Oses}}, \bibinfo {author}
  {\bibfnamefont {M.}~\bibnamefont {Samiee}}, \bibinfo {author} {\bibfnamefont
  {J.-P.}\ \bibnamefont {Maria}}, \bibinfo {author} {\bibfnamefont {D.~W.}\
  \bibnamefont {Brenner}}, \bibinfo {author} {\bibfnamefont {K.~S.}\
  \bibnamefont {Vecchio}},\ and\ \bibinfo {author} {\bibfnamefont
  {S.}~\bibnamefont {Curtarolo}},\ }\bibfield  {title} {\bibinfo {title}
  {High-entropy high-hardness metal carbides discovered by entropy
  descriptors},\ }\href@noop {} {\bibfield  {journal} {\bibinfo  {journal}
  {Nature communications}\ }\textbf {\bibinfo {volume} {9}},\ \bibinfo {pages}
  {1} (\bibinfo {year} {2018})}\BibitemShut {NoStop}%
\bibitem [{\citenamefont {Woods-Robinson}\ \emph {et~al.}(2022)\citenamefont
  {Woods-Robinson}, \citenamefont {Stevanovi{\'c}}, \citenamefont {Lany},
  \citenamefont {Heinselman}, \citenamefont {Horton}, \citenamefont {Persson},\
  and\ \citenamefont {Zakutayev}}]{woods2022role}%
  \BibitemOpen
  \bibfield  {author} {\bibinfo {author} {\bibfnamefont {R.}~\bibnamefont
  {Woods-Robinson}}, \bibinfo {author} {\bibfnamefont {V.}~\bibnamefont
  {Stevanovi{\'c}}}, \bibinfo {author} {\bibfnamefont {S.}~\bibnamefont
  {Lany}}, \bibinfo {author} {\bibfnamefont {K.~N.}\ \bibnamefont
  {Heinselman}}, \bibinfo {author} {\bibfnamefont {M.~K.}\ \bibnamefont
  {Horton}}, \bibinfo {author} {\bibfnamefont {K.~A.}\ \bibnamefont
  {Persson}},\ and\ \bibinfo {author} {\bibfnamefont {A.}~\bibnamefont
  {Zakutayev}},\ }\bibfield  {title} {\bibinfo {title} {Role of disorder in the
  synthesis of metastable zinc zirconium nitrides},\ }\href@noop {} {\bibfield
  {journal} {\bibinfo  {journal} {Physical Review Materials}\ }\textbf
  {\bibinfo {volume} {6}},\ \bibinfo {pages} {043804} (\bibinfo {year}
  {2022})}\BibitemShut {NoStop}%
\bibitem [{\citenamefont {Gorai}\ \emph {et~al.}(2016)\citenamefont {Gorai},
  \citenamefont {Toberer},\ and\ \citenamefont
  {Stevanovi{\'c}}}]{gorai2016thermoelectricity}%
  \BibitemOpen
  \bibfield  {author} {\bibinfo {author} {\bibfnamefont {P.}~\bibnamefont
  {Gorai}}, \bibinfo {author} {\bibfnamefont {E.~S.}\ \bibnamefont {Toberer}},\
  and\ \bibinfo {author} {\bibfnamefont {V.}~\bibnamefont {Stevanovi{\'c}}},\
  }\bibfield  {title} {\bibinfo {title} {Thermoelectricity in transition metal
  compounds: the role of spin disorder},\ }\href@noop {} {\bibfield  {journal}
  {\bibinfo  {journal} {Physical Chemistry Chemical Physics}\ }\textbf
  {\bibinfo {volume} {18}},\ \bibinfo {pages} {31777} (\bibinfo {year}
  {2016})}\BibitemShut {NoStop}%
\bibitem [{\citenamefont {Stevanovi{\'c}}(2016)}]{stevanovic2016sampling}%
  \BibitemOpen
  \bibfield  {author} {\bibinfo {author} {\bibfnamefont {V.}~\bibnamefont
  {Stevanovi{\'c}}},\ }\bibfield  {title} {\bibinfo {title} {Sampling
  polymorphs of ionic solids using random superlattices},\ }\href@noop {}
  {\bibfield  {journal} {\bibinfo  {journal} {Physical Review Letters}\
  }\textbf {\bibinfo {volume} {116}},\ \bibinfo {pages} {075503} (\bibinfo
  {year} {2016})}\BibitemShut {NoStop}%
\bibitem [{\citenamefont {Jones}\ and\ \citenamefont
  {Stevanovi{\'c}}(2017)}]{jones2017polymorphism}%
  \BibitemOpen
  \bibfield  {author} {\bibinfo {author} {\bibfnamefont {E.~B.}\ \bibnamefont
  {Jones}}\ and\ \bibinfo {author} {\bibfnamefont {V.}~\bibnamefont
  {Stevanovi{\'c}}},\ }\bibfield  {title} {\bibinfo {title} {Polymorphism in
  elemental silicon: Probabilistic interpretation of the realizability of
  metastable structures},\ }\href@noop {} {\bibfield  {journal} {\bibinfo
  {journal} {Physical Review B}\ }\textbf {\bibinfo {volume} {96}},\ \bibinfo
  {pages} {184101} (\bibinfo {year} {2017})}\BibitemShut {NoStop}%
\bibitem [{\citenamefont {Therrien}\ \emph {et~al.}(2021)\citenamefont
  {Therrien}, \citenamefont {Jones},\ and\ \citenamefont
  {Stevanovi{\'c}}}]{therrien2021metastable}%
  \BibitemOpen
  \bibfield  {author} {\bibinfo {author} {\bibfnamefont {F.}~\bibnamefont
  {Therrien}}, \bibinfo {author} {\bibfnamefont {E.~B.}\ \bibnamefont
  {Jones}},\ and\ \bibinfo {author} {\bibfnamefont {V.}~\bibnamefont
  {Stevanovi{\'c}}},\ }\bibfield  {title} {\bibinfo {title} {Metastable
  materials discovery in the age of large-scale computation},\ }\href@noop {}
  {\bibfield  {journal} {\bibinfo  {journal} {Applied Physics Reviews}\
  }\textbf {\bibinfo {volume} {8}},\ \bibinfo {pages} {031310} (\bibinfo {year}
  {2021})}\BibitemShut {NoStop}%
\bibitem [{\citenamefont {Jones}\ and\ \citenamefont
  {Stevanovi{\'c}}(2020)}]{jones2020glassy}%
  \BibitemOpen
  \bibfield  {author} {\bibinfo {author} {\bibfnamefont {E.~B.}\ \bibnamefont
  {Jones}}\ and\ \bibinfo {author} {\bibfnamefont {V.}~\bibnamefont
  {Stevanovi{\'c}}},\ }\bibfield  {title} {\bibinfo {title} {The glassy solid
  as a statistical ensemble of crystalline microstates},\ }\href@noop {}
  {\bibfield  {journal} {\bibinfo  {journal} {npj Computational Materials}\
  }\textbf {\bibinfo {volume} {6}},\ \bibinfo {pages} {1} (\bibinfo {year}
  {2020})}\BibitemShut {NoStop}%
\bibitem [{\citenamefont {Perim}\ \emph {et~al.}(2016)\citenamefont {Perim},
  \citenamefont {Lee}, \citenamefont {Liu}, \citenamefont {Toher},
  \citenamefont {Gong}, \citenamefont {Li}, \citenamefont {Simmons},
  \citenamefont {Levy}, \citenamefont {Vlassak}, \citenamefont {Schroers},\
  and\ \citenamefont {Curtarolo}}]{perim2016spectral}%
  \BibitemOpen
  \bibfield  {author} {\bibinfo {author} {\bibfnamefont {E.}~\bibnamefont
  {Perim}}, \bibinfo {author} {\bibfnamefont {D.}~\bibnamefont {Lee}}, \bibinfo
  {author} {\bibfnamefont {Y.}~\bibnamefont {Liu}}, \bibinfo {author}
  {\bibfnamefont {C.}~\bibnamefont {Toher}}, \bibinfo {author} {\bibfnamefont
  {P.}~\bibnamefont {Gong}}, \bibinfo {author} {\bibfnamefont {Y.}~\bibnamefont
  {Li}}, \bibinfo {author} {\bibfnamefont {W.~N.}\ \bibnamefont {Simmons}},
  \bibinfo {author} {\bibfnamefont {O.}~\bibnamefont {Levy}}, \bibinfo {author}
  {\bibfnamefont {J.~J.}\ \bibnamefont {Vlassak}}, \bibinfo {author}
  {\bibfnamefont {J.}~\bibnamefont {Schroers}},\ and\ \bibinfo {author}
  {\bibfnamefont {S.}~\bibnamefont {Curtarolo}},\ }\bibfield  {title} {\bibinfo
  {title} {Spectral descriptors for bulk metallic glasses based on the
  thermodynamics of competing crystalline phases},\ }\href@noop {} {\bibfield
  {journal} {\bibinfo  {journal} {Nature communications}\ }\textbf {\bibinfo
  {volume} {7}},\ \bibinfo {pages} {1} (\bibinfo {year} {2016})}\BibitemShut
  {NoStop}%
\bibitem [{\citenamefont {Kwak}\ and\ \citenamefont {Kim}(2017)}]{CLT}%
  \BibitemOpen
  \bibfield  {author} {\bibinfo {author} {\bibfnamefont {S.~G.}\ \bibnamefont
  {Kwak}}\ and\ \bibinfo {author} {\bibfnamefont {J.~H.}\ \bibnamefont {Kim}},\
  }\bibfield  {title} {\bibinfo {title} {Central limit theorem: the cornerstone
  of modern statistics},\ }\href@noop {} {\bibfield  {journal} {\bibinfo
  {journal} {Korean journal of anesthesiology}\ }\textbf {\bibinfo {volume}
  {70}},\ \bibinfo {pages} {144} (\bibinfo {year} {2017})}\BibitemShut
  {NoStop}%
\bibitem [{\citenamefont {Fischer}(2011)}]{CLT-History}%
  \BibitemOpen
  \bibfield  {author} {\bibinfo {author} {\bibfnamefont {H.}~\bibnamefont
  {Fischer}},\ }\href@noop {} {\emph {\bibinfo {title} {A history of the
  central limit theorem: from classical to modern probability theory}}}\
  (\bibinfo  {publisher} {Springer},\ \bibinfo {year} {2011})\BibitemShut
  {NoStop}%
\bibitem [{\citenamefont {Hart}\ and\ \citenamefont
  {Forcade}(2008)}]{hart2008algorithm}%
  \BibitemOpen
  \bibfield  {author} {\bibinfo {author} {\bibfnamefont {G.~L.}\ \bibnamefont
  {Hart}}\ and\ \bibinfo {author} {\bibfnamefont {R.~W.}\ \bibnamefont
  {Forcade}},\ }\bibfield  {title} {\bibinfo {title} {Algorithm for generating
  derivative structures},\ }\href@noop {} {\bibfield  {journal} {\bibinfo
  {journal} {Physical Review B}\ }\textbf {\bibinfo {volume} {77}},\ \bibinfo
  {pages} {224115} (\bibinfo {year} {2008})}\BibitemShut {NoStop}%
\bibitem [{\citenamefont {Kresse}\ and\ \citenamefont
  {Furthm\"{u}ller}(1996)}]{kresse_CMS:1996}%
  \BibitemOpen
  \bibfield  {author} {\bibinfo {author} {\bibfnamefont {G.}~\bibnamefont
  {Kresse}}\ and\ \bibinfo {author} {\bibfnamefont {J.}~\bibnamefont
  {Furthm\"{u}ller}},\ }\href@noop {} {\bibfield  {journal} {\bibinfo
  {journal} {Comput. Mater. Sci.}\ }\textbf {\bibinfo {volume} {6}},\ \bibinfo
  {pages} {15} (\bibinfo {year} {1996})}\BibitemShut {NoStop}%
\bibitem [{\citenamefont {d’Avezac}\ \emph {et~al.}(2010)\citenamefont
  {d’Avezac}, \citenamefont {Graf}, \citenamefont {Paudal}, \citenamefont
  {Peng}, \citenamefont {Zhang}, \citenamefont {Stephen},\ and\ \citenamefont
  {Stevanovic}}]{d2010pylada}%
  \BibitemOpen
  \bibfield  {author} {\bibinfo {author} {\bibfnamefont {M.}~\bibnamefont
  {d’Avezac}}, \bibinfo {author} {\bibfnamefont {P.}~\bibnamefont {Graf}},
  \bibinfo {author} {\bibfnamefont {T.}~\bibnamefont {Paudal}}, \bibinfo
  {author} {\bibfnamefont {H.}~\bibnamefont {Peng}}, \bibinfo {author}
  {\bibfnamefont {L.}~\bibnamefont {Zhang}}, \bibinfo {author} {\bibfnamefont
  {S.}~\bibnamefont {Stephen}},\ and\ \bibinfo {author} {\bibfnamefont
  {V.}~\bibnamefont {Stevanovic}},\ }\bibfield  {title} {\bibinfo {title}
  {Pylada: a comprehensive python framework for preparing, running, monitoring,
  analyzing, and archiving high throughput first principles calculations},\
  }\href@noop {} {\bibfield  {journal} {\bibinfo  {journal} {GitHub
  repository}\ } (\bibinfo {year} {2010})}\BibitemShut {NoStop}%
\bibitem [{\citenamefont {Esters}\ \emph {et~al.}(2021)\citenamefont {Esters},
  \citenamefont {Oses}, \citenamefont {Hicks}, \citenamefont {Mehl},
  \citenamefont {Jahn{\'a}tek}, \citenamefont {Hossain}, \citenamefont {Maria},
  \citenamefont {Brenner}, \citenamefont {Toher},\ and\ \citenamefont
  {Curtarolo}}]{esters2021settling}%
  \BibitemOpen
  \bibfield  {author} {\bibinfo {author} {\bibfnamefont {M.}~\bibnamefont
  {Esters}}, \bibinfo {author} {\bibfnamefont {C.}~\bibnamefont {Oses}},
  \bibinfo {author} {\bibfnamefont {D.}~\bibnamefont {Hicks}}, \bibinfo
  {author} {\bibfnamefont {M.~J.}\ \bibnamefont {Mehl}}, \bibinfo {author}
  {\bibfnamefont {M.}~\bibnamefont {Jahn{\'a}tek}}, \bibinfo {author}
  {\bibfnamefont {M.~D.}\ \bibnamefont {Hossain}}, \bibinfo {author}
  {\bibfnamefont {J.-P.}\ \bibnamefont {Maria}}, \bibinfo {author}
  {\bibfnamefont {D.~W.}\ \bibnamefont {Brenner}}, \bibinfo {author}
  {\bibfnamefont {C.}~\bibnamefont {Toher}},\ and\ \bibinfo {author}
  {\bibfnamefont {S.}~\bibnamefont {Curtarolo}},\ }\bibfield  {title} {\bibinfo
  {title} {Settling the matter of the role of vibrations in the stability of
  high-entropy carbides},\ }\href@noop {} {\bibfield  {journal} {\bibinfo
  {journal} {Nature communications}\ }\textbf {\bibinfo {volume} {12}},\
  \bibinfo {pages} {1} (\bibinfo {year} {2021})}\BibitemShut {NoStop}%
\bibitem [{\citenamefont {Islam}(2018)}]{islam2018sample}%
  \BibitemOpen
  \bibfield  {author} {\bibinfo {author} {\bibfnamefont {M.~R.}\ \bibnamefont
  {Islam}},\ }\bibfield  {title} {\bibinfo {title} {Sample size and its role in
  central limit theorem (clt)},\ }\href@noop {} {\bibfield  {journal} {\bibinfo
   {journal} {Computational and Applied Mathematics Journal}\ }\textbf
  {\bibinfo {volume} {4}},\ \bibinfo {pages} {1} (\bibinfo {year}
  {2018})}\BibitemShut {NoStop}%
\bibitem [{\citenamefont {Mendez}(1991)}]{mendez1991understanding}%
  \BibitemOpen
  \bibfield  {author} {\bibinfo {author} {\bibfnamefont {H.}~\bibnamefont
  {Mendez}},\ }\href@noop {} {\emph {\bibinfo {title} {Understanding the
  central limit theorem}}}\ (\bibinfo  {publisher} {University of California,
  Santa Barbara},\ \bibinfo {year} {1991})\BibitemShut {NoStop}%
\bibitem [{\citenamefont {Perdew}\ \emph {et~al.}(1996)\citenamefont {Perdew},
  \citenamefont {Burke},\ and\ \citenamefont {Ernzerhof}}]{perdew_PRL:1996}%
  \BibitemOpen
  \bibfield  {author} {\bibinfo {author} {\bibfnamefont {J.~P.}\ \bibnamefont
  {Perdew}}, \bibinfo {author} {\bibfnamefont {K.}~\bibnamefont {Burke}},\ and\
  \bibinfo {author} {\bibfnamefont {M.}~\bibnamefont {Ernzerhof}},\ }\bibfield
  {title} {\bibinfo {title} {Generalized gradient approximation made simple},\
  }\href {https://doi.org/10.1103/PhysRevLett.77.3865} {\bibfield  {journal}
  {\bibinfo  {journal} {Phys. Rev. Lett.}\ }\textbf {\bibinfo {volume} {77}},\
  \bibinfo {pages} {3865} (\bibinfo {year} {1996})}\BibitemShut {NoStop}%
\bibitem [{\citenamefont {Bl\"ochl}(1994)}]{bloechl_PRB:1994}%
  \BibitemOpen
  \bibfield  {author} {\bibinfo {author} {\bibfnamefont {P.~E.}\ \bibnamefont
  {Bl\"ochl}},\ }\bibfield  {title} {\bibinfo {title} {Projector augmented-wave
  method},\ }\href {https://doi.org/10.1103/PhysRevB.50.17953} {\bibfield
  {journal} {\bibinfo  {journal} {Phys. Rev. B}\ }\textbf {\bibinfo {volume}
  {50}},\ \bibinfo {pages} {17953} (\bibinfo {year} {1994})}\BibitemShut
  {NoStop}%
\bibitem [{\citenamefont {Van~de Walle}\ \emph {et~al.}(2013)\citenamefont
  {Van~de Walle}, \citenamefont {Tiwary}, \citenamefont {De~Jong},
  \citenamefont {Olmsted}, \citenamefont {Asta}, \citenamefont {Dick},
  \citenamefont {Shin}, \citenamefont {Wang}, \citenamefont {Chen},\ and\
  \citenamefont {Liu}}]{mcsqs}%
  \BibitemOpen
  \bibfield  {author} {\bibinfo {author} {\bibfnamefont {A.}~\bibnamefont
  {Van~de Walle}}, \bibinfo {author} {\bibfnamefont {P.}~\bibnamefont
  {Tiwary}}, \bibinfo {author} {\bibfnamefont {M.}~\bibnamefont {De~Jong}},
  \bibinfo {author} {\bibfnamefont {D.}~\bibnamefont {Olmsted}}, \bibinfo
  {author} {\bibfnamefont {M.}~\bibnamefont {Asta}}, \bibinfo {author}
  {\bibfnamefont {A.}~\bibnamefont {Dick}}, \bibinfo {author} {\bibfnamefont
  {D.}~\bibnamefont {Shin}}, \bibinfo {author} {\bibfnamefont {Y.}~\bibnamefont
  {Wang}}, \bibinfo {author} {\bibfnamefont {L.-Q.}\ \bibnamefont {Chen}},\
  and\ \bibinfo {author} {\bibfnamefont {Z.-K.}\ \bibnamefont {Liu}},\
  }\bibfield  {title} {\bibinfo {title} {Efficient stochastic generation of
  special quasirandom structures},\ }\href@noop {} {\bibfield  {journal}
  {\bibinfo  {journal} {Calphad}\ }\textbf {\bibinfo {volume} {42}},\ \bibinfo
  {pages} {13} (\bibinfo {year} {2013})}\BibitemShut {NoStop}%
\bibitem [{\citenamefont {Birch}(1947)}]{birch1947finite}%
  \BibitemOpen
  \bibfield  {author} {\bibinfo {author} {\bibfnamefont {F.}~\bibnamefont
  {Birch}},\ }\bibfield  {title} {\bibinfo {title} {Finite elastic strain of
  cubic crystals},\ }\href@noop {} {\bibfield  {journal} {\bibinfo  {journal}
  {Physical review}\ }\textbf {\bibinfo {volume} {71}},\ \bibinfo {pages} {809}
  (\bibinfo {year} {1947})}\BibitemShut {NoStop}%
\bibitem [{\citenamefont {Murnaghan}(1944)}]{murnaghan1944compressibility}%
  \BibitemOpen
  \bibfield  {author} {\bibinfo {author} {\bibfnamefont {F.~D.}\ \bibnamefont
  {Murnaghan}},\ }\bibfield  {title} {\bibinfo {title} {The compressibility of
  media under extreme pressures},\ }\href@noop {} {\bibfield  {journal}
  {\bibinfo  {journal} {Proceedings of the National Academy of Sciences}\
  }\textbf {\bibinfo {volume} {30}},\ \bibinfo {pages} {244} (\bibinfo {year}
  {1944})}\BibitemShut {NoStop}%
\bibitem [{\citenamefont {Reu{\ss}}(1929)}]{reuss1929berechnung}%
  \BibitemOpen
  \bibfield  {author} {\bibinfo {author} {\bibfnamefont {A.}~\bibnamefont
  {Reu{\ss}}},\ }\bibfield  {title} {\bibinfo {title} {Berechnung der
  flie{\ss}grenze von mischkristallen auf grund der plastizit{\"a}tsbedingung
  f{\"u}r einkristalle.},\ }\href@noop {} {\bibfield  {journal} {\bibinfo
  {journal} {ZAMM-Journal of Applied Mathematics and Mechanics/Zeitschrift
  f{\"u}r Angewandte Mathematik und Mechanik}\ }\textbf {\bibinfo {volume}
  {9}},\ \bibinfo {pages} {49} (\bibinfo {year} {1929})}\BibitemShut {NoStop}%
\bibitem [{\citenamefont {Peterson}\ \emph {et~al.}(2001)\citenamefont
  {Peterson}, \citenamefont {Proffen}, \citenamefont {Jeong}, \citenamefont
  {Billinge}, \citenamefont {Choi}, \citenamefont {Kanatzidis},\ and\
  \citenamefont {Radaelli}}]{peterson2001local}%
  \BibitemOpen
  \bibfield  {author} {\bibinfo {author} {\bibfnamefont {P.~F.}\ \bibnamefont
  {Peterson}}, \bibinfo {author} {\bibfnamefont {T.}~\bibnamefont {Proffen}},
  \bibinfo {author} {\bibfnamefont {I.-K.}\ \bibnamefont {Jeong}}, \bibinfo
  {author} {\bibfnamefont {S.~J.}\ \bibnamefont {Billinge}}, \bibinfo {author}
  {\bibfnamefont {K.-S.}\ \bibnamefont {Choi}}, \bibinfo {author}
  {\bibfnamefont {M.~G.}\ \bibnamefont {Kanatzidis}},\ and\ \bibinfo {author}
  {\bibfnamefont {P.~G.}\ \bibnamefont {Radaelli}},\ }\bibfield  {title}
  {\bibinfo {title} {Local atomic strain in znse 1- x te x from high real-space
  resolution neutron pair distribution function measurements},\ }\href@noop {}
  {\bibfield  {journal} {\bibinfo  {journal} {Physical Review B}\ }\textbf
  {\bibinfo {volume} {63}},\ \bibinfo {pages} {165211} (\bibinfo {year}
  {2001})}\BibitemShut {NoStop}%
\bibitem [{\citenamefont {Doak}\ and\ \citenamefont
  {Wolverton}(2012)}]{doak2012coherent}%
  \BibitemOpen
  \bibfield  {author} {\bibinfo {author} {\bibfnamefont {J.~W.}\ \bibnamefont
  {Doak}}\ and\ \bibinfo {author} {\bibfnamefont {C.}~\bibnamefont
  {Wolverton}},\ }\bibfield  {title} {\bibinfo {title} {Coherent and incoherent
  phase stabilities of thermoelectric rocksalt iv-vi semiconductor alloys},\
  }\href@noop {} {\bibfield  {journal} {\bibinfo  {journal} {Physical Review
  B}\ }\textbf {\bibinfo {volume} {86}},\ \bibinfo {pages} {144202} (\bibinfo
  {year} {2012})}\BibitemShut {NoStop}%
\bibitem [{\citenamefont {Ortiz}\ \emph {et~al.}(2019)\citenamefont {Ortiz},
  \citenamefont {Adamczyk}, \citenamefont {Gordiz}, \citenamefont {Braden},\
  and\ \citenamefont {Toberer}}]{ortiz2019towards}%
  \BibitemOpen
  \bibfield  {author} {\bibinfo {author} {\bibfnamefont {B.~R.}\ \bibnamefont
  {Ortiz}}, \bibinfo {author} {\bibfnamefont {J.~M.}\ \bibnamefont {Adamczyk}},
  \bibinfo {author} {\bibfnamefont {K.}~\bibnamefont {Gordiz}}, \bibinfo
  {author} {\bibfnamefont {T.}~\bibnamefont {Braden}},\ and\ \bibinfo {author}
  {\bibfnamefont {E.~S.}\ \bibnamefont {Toberer}},\ }\bibfield  {title}
  {\bibinfo {title} {Towards the high-throughput synthesis of bulk materials:
  thermoelectric pbte--pbse--snte--snse alloys},\ }\href@noop {} {\bibfield
  {journal} {\bibinfo  {journal} {Molecular Systems Design \& Engineering}\
  }\textbf {\bibinfo {volume} {4}},\ \bibinfo {pages} {407} (\bibinfo {year}
  {2019})}\BibitemShut {NoStop}%
\bibitem [{\citenamefont {Liu}\ and\ \citenamefont
  {Chang}(1994)}]{PbSSeTe-Phase-Diagram}%
  \BibitemOpen
  \bibfield  {author} {\bibinfo {author} {\bibfnamefont {H.}~\bibnamefont
  {Liu}}\ and\ \bibinfo {author} {\bibfnamefont {L.}~\bibnamefont {Chang}},\
  }\bibfield  {title} {\bibinfo {title} {Phase relations in the system
  pbs-pbse-pbte},\ }\href@noop {} {\bibfield  {journal} {\bibinfo  {journal}
  {Mineralogical Magazine}\ }\textbf {\bibinfo {volume} {58}},\ \bibinfo
  {pages} {567} (\bibinfo {year} {1994})}\BibitemShut {NoStop}%
\bibitem [{\citenamefont {Zhang}\ \emph {et~al.}(2012)\citenamefont {Zhang},
  \citenamefont {Cao}, \citenamefont {Liu}, \citenamefont {Lukas},
  \citenamefont {Yu}, \citenamefont {Chen}, \citenamefont {Opeil},
  \citenamefont {Broido}, \citenamefont {Chen},\ and\ \citenamefont
  {Ren}}]{zhang2012heavy}%
  \BibitemOpen
  \bibfield  {author} {\bibinfo {author} {\bibfnamefont {Q.}~\bibnamefont
  {Zhang}}, \bibinfo {author} {\bibfnamefont {F.}~\bibnamefont {Cao}}, \bibinfo
  {author} {\bibfnamefont {W.}~\bibnamefont {Liu}}, \bibinfo {author}
  {\bibfnamefont {K.}~\bibnamefont {Lukas}}, \bibinfo {author} {\bibfnamefont
  {B.}~\bibnamefont {Yu}}, \bibinfo {author} {\bibfnamefont {S.}~\bibnamefont
  {Chen}}, \bibinfo {author} {\bibfnamefont {C.}~\bibnamefont {Opeil}},
  \bibinfo {author} {\bibfnamefont {D.}~\bibnamefont {Broido}}, \bibinfo
  {author} {\bibfnamefont {G.}~\bibnamefont {Chen}},\ and\ \bibinfo {author}
  {\bibfnamefont {Z.}~\bibnamefont {Ren}},\ }\bibfield  {title} {\bibinfo
  {title} {Heavy doping and band engineering by potassium to improve the
  thermoelectric figure of merit in p-type pbte, pbse, and pbte1--y se y},\
  }\href@noop {} {\bibfield  {journal} {\bibinfo  {journal} {Journal of the
  American chemical society}\ }\textbf {\bibinfo {volume} {134}},\ \bibinfo
  {pages} {10031} (\bibinfo {year} {2012})}\BibitemShut {NoStop}%
\bibitem [{\citenamefont {Hume-Rothery}\ and\ \citenamefont
  {Powell}(1935)}]{hume1935theory}%
  \BibitemOpen
  \bibfield  {author} {\bibinfo {author} {\bibfnamefont {W.}~\bibnamefont
  {Hume-Rothery}}\ and\ \bibinfo {author} {\bibfnamefont {H.~M.}\ \bibnamefont
  {Powell}},\ }\bibfield  {title} {\bibinfo {title} {On the theory of
  super-lattice structures in alloys},\ }\href@noop {} {\bibfield  {journal}
  {\bibinfo  {journal} {Zeitschrift f{\"u}r Kristallographie-Crystalline
  Materials}\ }\textbf {\bibinfo {volume} {91}},\ \bibinfo {pages} {23}
  (\bibinfo {year} {1935})}\BibitemShut {NoStop}%
\bibitem [{\citenamefont {Usanmaz}\ \emph {et~al.}(2016)\citenamefont
  {Usanmaz}, \citenamefont {Nath}, \citenamefont {Plata}, \citenamefont {Hart},
  \citenamefont {Takeuchi}, \citenamefont {Nardelli}, \citenamefont {Fornari},\
  and\ \citenamefont {Curtarolo}}]{usanmaz2016first}%
  \BibitemOpen
  \bibfield  {author} {\bibinfo {author} {\bibfnamefont {D.}~\bibnamefont
  {Usanmaz}}, \bibinfo {author} {\bibfnamefont {P.}~\bibnamefont {Nath}},
  \bibinfo {author} {\bibfnamefont {J.~J.}\ \bibnamefont {Plata}}, \bibinfo
  {author} {\bibfnamefont {G.~L.}\ \bibnamefont {Hart}}, \bibinfo {author}
  {\bibfnamefont {I.}~\bibnamefont {Takeuchi}}, \bibinfo {author}
  {\bibfnamefont {M.~B.}\ \bibnamefont {Nardelli}}, \bibinfo {author}
  {\bibfnamefont {M.}~\bibnamefont {Fornari}},\ and\ \bibinfo {author}
  {\bibfnamefont {S.}~\bibnamefont {Curtarolo}},\ }\bibfield  {title} {\bibinfo
  {title} {First principles thermodynamical modeling of the binodal and
  spinodal curves in lead chalcogenides},\ }\href@noop {} {\bibfield  {journal}
  {\bibinfo  {journal} {Physical Chemistry Chemical Physics}\ }\textbf
  {\bibinfo {volume} {18}},\ \bibinfo {pages} {5005} (\bibinfo {year}
  {2016})}\BibitemShut {NoStop}%
\bibitem [{\citenamefont {Shannon}(1976)}]{shannon1976revised}%
  \BibitemOpen
  \bibfield  {author} {\bibinfo {author} {\bibfnamefont {R.~D.}\ \bibnamefont
  {Shannon}},\ }\bibfield  {title} {\bibinfo {title} {Revised effective ionic
  radii and systematic studies of interatomic distances in halides and
  chalcogenides},\ }\href@noop {} {\bibfield  {journal} {\bibinfo  {journal}
  {Acta crystallographica section A: crystal physics, diffraction, theoretical
  and general crystallography}\ }\textbf {\bibinfo {volume} {32}},\ \bibinfo
  {pages} {751} (\bibinfo {year} {1976})}\BibitemShut {NoStop}%
\bibitem [{\citenamefont {Bissert}\ and\ \citenamefont {Hesse}(1978)}]{GeS}%
  \BibitemOpen
  \bibfield  {author} {\bibinfo {author} {\bibfnamefont {G.}~\bibnamefont
  {Bissert}}\ and\ \bibinfo {author} {\bibfnamefont {K.-F.}\ \bibnamefont
  {Hesse}},\ }\bibfield  {title} {\bibinfo {title} {Verfeinerung der struktur
  von germanium (ii)-sulfid, ges},\ }\href@noop {} {\bibfield  {journal}
  {\bibinfo  {journal} {Acta Crystallographica Section B: Structural
  Crystallography and Crystal Chemistry}\ }\textbf {\bibinfo {volume} {34}},\
  \bibinfo {pages} {1322} (\bibinfo {year} {1978})}\BibitemShut {NoStop}%
\bibitem [{\citenamefont {Murgatroyd}\ \emph {et~al.}(2020)\citenamefont
  {Murgatroyd}, \citenamefont {Smiles}, \citenamefont {Savory}, \citenamefont
  {Shalvey}, \citenamefont {Swallow}, \citenamefont {Fleck}, \citenamefont
  {Robertson}, \citenamefont {Jackel}, \citenamefont {Alaria}, \citenamefont
  {Major}, \citenamefont {Scanlon},\ and\ \citenamefont {Veal}}]{GeSe}%
  \BibitemOpen
  \bibfield  {author} {\bibinfo {author} {\bibfnamefont {P.~A.}\ \bibnamefont
  {Murgatroyd}}, \bibinfo {author} {\bibfnamefont {M.~J.}\ \bibnamefont
  {Smiles}}, \bibinfo {author} {\bibfnamefont {C.~N.}\ \bibnamefont {Savory}},
  \bibinfo {author} {\bibfnamefont {T.~P.}\ \bibnamefont {Shalvey}}, \bibinfo
  {author} {\bibfnamefont {J.~E.}\ \bibnamefont {Swallow}}, \bibinfo {author}
  {\bibfnamefont {N.}~\bibnamefont {Fleck}}, \bibinfo {author} {\bibfnamefont
  {C.~M.}\ \bibnamefont {Robertson}}, \bibinfo {author} {\bibfnamefont
  {F.}~\bibnamefont {Jackel}}, \bibinfo {author} {\bibfnamefont
  {J.}~\bibnamefont {Alaria}}, \bibinfo {author} {\bibfnamefont {J.~D.}\
  \bibnamefont {Major}}, \bibinfo {author} {\bibfnamefont {D.~O.}\ \bibnamefont
  {Scanlon}},\ and\ \bibinfo {author} {\bibfnamefont {T.~D.}\ \bibnamefont
  {Veal}},\ }\bibfield  {title} {\bibinfo {title} {Gese: optical spectroscopy
  and theoretical study of a van der waals solar absorber},\ }\href@noop {}
  {\bibfield  {journal} {\bibinfo  {journal} {Chemistry of Materials}\ }\textbf
  {\bibinfo {volume} {32}},\ \bibinfo {pages} {3245} (\bibinfo {year}
  {2020})}\BibitemShut {NoStop}%
\bibitem [{\citenamefont {Samanta}\ \emph {et~al.}(2019)\citenamefont
  {Samanta}, \citenamefont {Ghosh}, \citenamefont {Arora}, \citenamefont
  {Waghmare},\ and\ \citenamefont {Biswas}}]{GeTe}%
  \BibitemOpen
  \bibfield  {author} {\bibinfo {author} {\bibfnamefont {M.}~\bibnamefont
  {Samanta}}, \bibinfo {author} {\bibfnamefont {T.}~\bibnamefont {Ghosh}},
  \bibinfo {author} {\bibfnamefont {R.}~\bibnamefont {Arora}}, \bibinfo
  {author} {\bibfnamefont {U.~V.}\ \bibnamefont {Waghmare}},\ and\ \bibinfo
  {author} {\bibfnamefont {K.}~\bibnamefont {Biswas}},\ }\bibfield  {title}
  {\bibinfo {title} {Realization of both n-and p-type gete thermoelectrics:
  electronic structure modulation by agbise2 alloying},\ }\href@noop {}
  {\bibfield  {journal} {\bibinfo  {journal} {Journal of the American Chemical
  Society}\ }\textbf {\bibinfo {volume} {141}},\ \bibinfo {pages} {19505}
  (\bibinfo {year} {2019})}\BibitemShut {NoStop}%
\bibitem [{\citenamefont {Perumal}\ \emph {et~al.}(2016)\citenamefont
  {Perumal}, \citenamefont {Roychowdhury},\ and\ \citenamefont
  {Biswas}}]{GeTe-Thermoelectrics}%
  \BibitemOpen
  \bibfield  {author} {\bibinfo {author} {\bibfnamefont {S.}~\bibnamefont
  {Perumal}}, \bibinfo {author} {\bibfnamefont {S.}~\bibnamefont
  {Roychowdhury}},\ and\ \bibinfo {author} {\bibfnamefont {K.}~\bibnamefont
  {Biswas}},\ }\bibfield  {title} {\bibinfo {title} {High performance
  thermoelectric materials and devices based on gete},\ }\href@noop {}
  {\bibfield  {journal} {\bibinfo  {journal} {Journal of Materials Chemistry
  C}\ }\textbf {\bibinfo {volume} {4}},\ \bibinfo {pages} {7520} (\bibinfo
  {year} {2016})}\BibitemShut {NoStop}%
\bibitem [{\citenamefont {Samanta}\ and\ \citenamefont
  {Biswas}(2017)}]{GeTeSeS-Thermoelectrics}%
  \BibitemOpen
  \bibfield  {author} {\bibinfo {author} {\bibfnamefont {M.}~\bibnamefont
  {Samanta}}\ and\ \bibinfo {author} {\bibfnamefont {K.}~\bibnamefont
  {Biswas}},\ }\bibfield  {title} {\bibinfo {title} {Low thermal conductivity
  and high thermoelectric performance in (gete) 1--2 x (gese) x (ges) x:
  competition between solid solution and phase separation},\ }\href@noop {}
  {\bibfield  {journal} {\bibinfo  {journal} {Journal of the American Chemical
  Society}\ }\textbf {\bibinfo {volume} {139}},\ \bibinfo {pages} {9382}
  (\bibinfo {year} {2017})}\BibitemShut {NoStop}%
\bibitem [{\citenamefont {Adamczyk}\ \emph {et~al.}(2022)\citenamefont
  {Adamczyk}, \citenamefont {Bipasha}, \citenamefont {Rome}, \citenamefont
  {Ciesielski}, \citenamefont {Ertekin},\ and\ \citenamefont {Toberer}}]{MGT}%
  \BibitemOpen
  \bibfield  {author} {\bibinfo {author} {\bibfnamefont {J.~M.}\ \bibnamefont
  {Adamczyk}}, \bibinfo {author} {\bibfnamefont {F.~A.}\ \bibnamefont
  {Bipasha}}, \bibinfo {author} {\bibfnamefont {G.~A.}\ \bibnamefont {Rome}},
  \bibinfo {author} {\bibfnamefont {K.}~\bibnamefont {Ciesielski}}, \bibinfo
  {author} {\bibfnamefont {E.}~\bibnamefont {Ertekin}},\ and\ \bibinfo {author}
  {\bibfnamefont {E.~S.}\ \bibnamefont {Toberer}},\ }\bibfield  {title}
  {\bibinfo {title} {Symmetry breaking in ge 1- x mn x te and the impact on
  thermoelectric transport},\ }\href@noop {} {\bibfield  {journal} {\bibinfo
  {journal} {Journal of Materials Chemistry A}\ }\textbf {\bibinfo {volume}
  {10}},\ \bibinfo {pages} {16468} (\bibinfo {year} {2022})}\BibitemShut
  {NoStop}%
\bibitem [{\citenamefont {Villars}\ \emph {et~al.}(2006)\citenamefont
  {Villars}, \citenamefont {Okamoto},\ and\ \citenamefont {Cenzual}}]{GeTeS}%
  \BibitemOpen
  \bibfield  {author} {\bibinfo {author} {\bibfnamefont {P.}~\bibnamefont
  {Villars}}, \bibinfo {author} {\bibfnamefont {H.}~\bibnamefont {Okamoto}},\
  and\ \bibinfo {author} {\bibfnamefont {K.}~\bibnamefont {Cenzual}},\
  }\bibfield  {title} {\bibinfo {title} {Asm alloy phase diagrams database},\
  }\href@noop {} {\bibfield  {journal} {\bibinfo  {journal} {ASM International,
  Materials Park, OH, USA}\ } (\bibinfo {year} {2006})}\BibitemShut {NoStop}%
\bibitem [{\citenamefont {Koren}\ \emph {et~al.}(1983)\citenamefont {Koren},
  \citenamefont {Kindyak},\ and\ \citenamefont {Matyas}}]{GeSeS}%
  \BibitemOpen
  \bibfield  {author} {\bibinfo {author} {\bibfnamefont {N.}~\bibnamefont
  {Koren}}, \bibinfo {author} {\bibfnamefont {V.}~\bibnamefont {Kindyak}},\
  and\ \bibinfo {author} {\bibfnamefont {E.}~\bibnamefont {Matyas}},\
  }\bibfield  {title} {\bibinfo {title} {Phase diagram of the system
  ges-gese},\ }\href@noop {} {\bibfield  {journal} {\bibinfo  {journal}
  {physica status solidi (a)}\ }\textbf {\bibinfo {volume} {80}},\ \bibinfo
  {pages} {K105} (\bibinfo {year} {1983})}\BibitemShut {NoStop}%
\bibitem [{\citenamefont {Wiedemeier}\ and\ \citenamefont
  {Siemers}(1984)}]{GeTeSe}%
  \BibitemOpen
  \bibfield  {author} {\bibinfo {author} {\bibfnamefont {H.}~\bibnamefont
  {Wiedemeier}}\ and\ \bibinfo {author} {\bibfnamefont {P.~A.}\ \bibnamefont
  {Siemers}},\ }\bibfield  {title} {\bibinfo {title} {The
  temperature—composition phase diagram of the gese—gete system},\ }in\
  \href@noop {} {\emph {\bibinfo {booktitle} {Modern High Temperature
  Science}}}\ (\bibinfo  {publisher} {Springer},\ \bibinfo {year} {1984})\ pp.\
  \bibinfo {pages} {395--408}\BibitemShut {NoStop}%
\bibitem [{\citenamefont {Magunov}\ \emph {et~al.}(1992)\citenamefont
  {Magunov}, \citenamefont {Belyuga},\ and\ \citenamefont {Magunov}}]{PbGeS}%
  \BibitemOpen
  \bibfield  {author} {\bibinfo {author} {\bibfnamefont {R.}~\bibnamefont
  {Magunov}}, \bibinfo {author} {\bibfnamefont {Y.~V.}\ \bibnamefont
  {Belyuga}},\ and\ \bibinfo {author} {\bibfnamefont {I.}~\bibnamefont
  {Magunov}},\ }\bibfield  {title} {\bibinfo {title} {The pbs-ges system},\
  }\href@noop {} {\bibfield  {journal} {\bibinfo  {journal} {Russian journal of
  inorganic chemistry}\ }\textbf {\bibinfo {volume} {37}},\ \bibinfo {pages}
  {1345} (\bibinfo {year} {1992})}\BibitemShut {NoStop}%
\bibitem [{\citenamefont {Shelimova}\ \emph {et~al.}(1966)\citenamefont
  {Shelimova}, \citenamefont {Abrikosov}, \citenamefont {Zhdanova},\ and\
  \citenamefont {Sizov}}]{PbGeSe}%
  \BibitemOpen
  \bibfield  {author} {\bibinfo {author} {\bibfnamefont {L.}~\bibnamefont
  {Shelimova}}, \bibinfo {author} {\bibfnamefont {N.}~\bibnamefont
  {Abrikosov}}, \bibinfo {author} {\bibfnamefont {V.}~\bibnamefont
  {Zhdanova}},\ and\ \bibinfo {author} {\bibfnamefont {V.}~\bibnamefont
  {Sizov}},\ }\bibfield  {title} {\bibinfo {title} {Investigation of the
  pbse-gese and gese-gete systems},\ }\href@noop {} {\bibfield  {journal}
  {\bibinfo  {journal} {Izv Akad Nauk SSSR Neorgan Materialy}\ }\textbf
  {\bibinfo {volume} {2}},\ \bibinfo {pages} {2103} (\bibinfo {year}
  {1966})}\BibitemShut {NoStop}%
\bibitem [{\citenamefont {Yashina}\ and\ \citenamefont {Leute}(2000)}]{PbGeTe}%
  \BibitemOpen
  \bibfield  {author} {\bibinfo {author} {\bibfnamefont {L.}~\bibnamefont
  {Yashina}}\ and\ \bibinfo {author} {\bibfnamefont {V.}~\bibnamefont
  {Leute}},\ }\bibfield  {title} {\bibinfo {title} {The phase diagrams of the
  quasibinary systems (pb, ge) te and (ge, sn) te},\ }\href@noop {} {\bibfield
  {journal} {\bibinfo  {journal} {Journal of alloys and compounds}\ }\textbf
  {\bibinfo {volume} {313}},\ \bibinfo {pages} {85} (\bibinfo {year}
  {2000})}\BibitemShut {NoStop}%
\bibitem [{\citenamefont {Settles}(2009)}]{settles2009active}%
  \BibitemOpen
  \bibfield  {author} {\bibinfo {author} {\bibfnamefont {B.}~\bibnamefont
  {Settles}},\ }\bibfield  {title} {\bibinfo {title} {Active learning
  literature survey},\ }\href@noop {} {\  (\bibinfo {year} {2009})}\BibitemShut
  {NoStop}%
\bibitem [{\citenamefont {Garnett}(2022)}]{garnett2022bayesian}%
  \BibitemOpen
  \bibfield  {author} {\bibinfo {author} {\bibfnamefont {R.}~\bibnamefont
  {Garnett}},\ }\href@noop {} {\bibinfo {title} {Bayesian optimization}}
  (\bibinfo {year} {2022})\BibitemShut {NoStop}%
\bibitem [{\citenamefont {Tokdar}\ and\ \citenamefont
  {Kass}(2010)}]{importance}%
  \BibitemOpen
  \bibfield  {author} {\bibinfo {author} {\bibfnamefont {S.~T.}\ \bibnamefont
  {Tokdar}}\ and\ \bibinfo {author} {\bibfnamefont {R.~E.}\ \bibnamefont
  {Kass}},\ }\bibfield  {title} {\bibinfo {title} {Importance sampling: a
  review},\ }\href@noop {} {\bibfield  {journal} {\bibinfo  {journal} {Wiley
  Interdisciplinary Reviews: Computational Statistics}\ }\textbf {\bibinfo
  {volume} {2}},\ \bibinfo {pages} {54} (\bibinfo {year} {2010})}\BibitemShut
  {NoStop}%
\bibitem [{\citenamefont {Wallace}\ \emph {et~al.}(2019)\citenamefont
  {Wallace}, \citenamefont {Frost},\ and\ \citenamefont
  {Walsh}}]{wallace2019atomistic}%
  \BibitemOpen
  \bibfield  {author} {\bibinfo {author} {\bibfnamefont {S.~K.}\ \bibnamefont
  {Wallace}}, \bibinfo {author} {\bibfnamefont {J.~M.}\ \bibnamefont {Frost}},\
  and\ \bibinfo {author} {\bibfnamefont {A.}~\bibnamefont {Walsh}},\ }\bibfield
   {title} {\bibinfo {title} {Atomistic insights into the order--disorder
  transition in cu 2 znsns 4 solar cells from monte carlo simulations},\
  }\href@noop {} {\bibfield  {journal} {\bibinfo  {journal} {Journal of
  Materials Chemistry A}\ }\textbf {\bibinfo {volume} {7}},\ \bibinfo {pages}
  {312} (\bibinfo {year} {2019})}\BibitemShut {NoStop}%
\end{thebibliography}%

\end{document}


\title{Supplemental Materials}
\maketitle

\section{PbSe$_{1-x}$Te$_{x}$ Bonds}
\begin{figure}[h!]
  \includegraphics[width = 0.9 \linewidth]{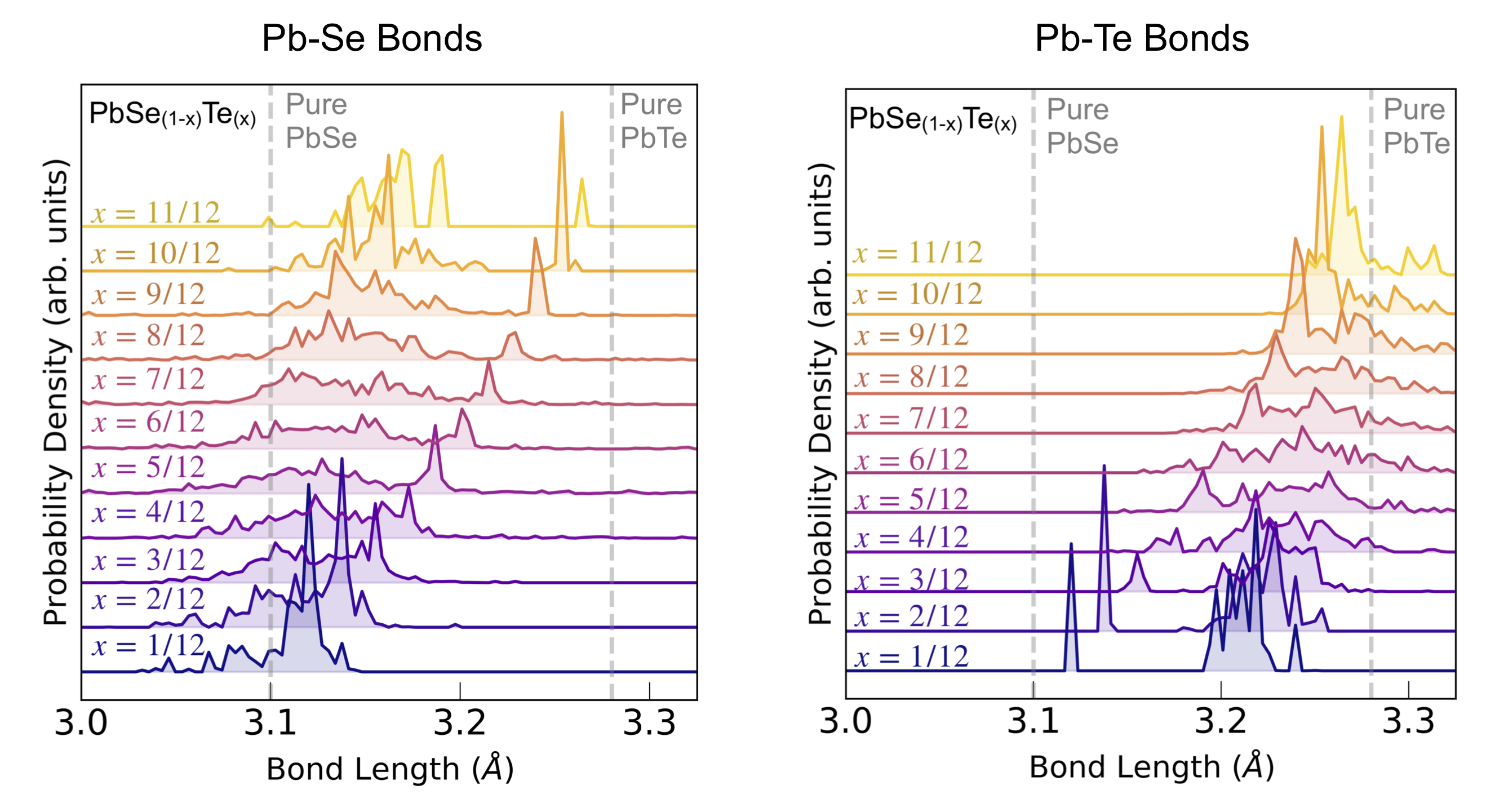}
  \caption{Pb-Se bonds are shorter than the Vegard's law bond length, while Pb-Te bond lengths are longer. The grey dashed lines show the bond lengths of pure PbSe and PbTe. The Vegard's law bond length is thus a linear interpolation between those two extrema. Each distribution is normalized such that their integrals are equivalent.}
\end{figure}

\section{Implementing the CLT at finite temperatures}
For property X, the ensemble average of the property, $\tilde{X}$ can be calculated at finite temperatures. Caution needs to be taken since we are not sampling from the true distribution of configurations, but rather we are using a uniform distribution as a proxy. Here we use $\tilde{X}$ to denote that the ensemble average is no longer an arithmetic average, like $\bar{X}$. We start with the same expression used in the main text:
\begin{equation}
    \tilde{X}=\sum_i^n P_i*X_i.
\end{equation}
This can be expanded out:
\begin{equation}
    \tilde{X}=\frac{\sum_i^n e^{-E_i/k_BT}X_i}{\sum_i^n e^{-E_i/k_BT}}.
\end{equation}
The above gives us the form used by Tokdar et al. From Tokdar et al.~, the variance of the distribution is:
\begin{equation}
    \sigma_{X,n}^2=\frac{\sum_i^n (e^{-E_i/k_BT})^2 (X_i-\bar{X})^2}{[\sum_i^n (e^{-E_i/k_BT})]^2}.
\end{equation}
Finally, the above expression can be simplified and used to derive the standard deviation, $\sigma_{\tilde{X},n}$:
\begin{equation}
\begin{aligned}
    \sigma_{\tilde{X},n}^2=&\frac{\sum_i^n (e^{-E_i/k_BT})^2 (X_i-\bar{X})^2}{Z^2}\\
    \sigma_{\tilde{X},n}^2=&\sum_i^n (\frac{e^{-E_i/k_BT}}{Z})^2 (X_i-\bar{X})^2\\
    \sigma_{\tilde{X},n}^2=&\sum_i^n P_i^2 (X_i-\bar{X})^2\\
    \sigma_{\tilde{X},n}=&\sqrt{\sum_i^n P_i^2 (X_i-\bar{X})^2}
\end{aligned}
\end{equation}

As a sanity check, we can show that the above equation reduces down to eq.~[9] when sampling from the true distribution. When sampling from the true distribution, no reweighting of the configurations is required. Thus, the probabilities are all equal ($n^{-1}$), and the ensemble average reduces down to an arithmetic average. Substituting this in:
\begin{equation}
\begin{aligned}
   \sigma_{\bar{X},n}=&\sqrt{\frac{1}{n^2} \sum_i^n (X_i-\bar{X})^2}\\
   \sigma_{\bar{X},n}=&\frac{1}{\sqrt{n}} \sqrt{\sum_i^n \frac{(X_i-\bar{X})^2}{n}}\\
   \sigma_{\bar{X},n}=&\frac{\sigma_{X,n}}{\sqrt{n}}
\end{aligned}
\end{equation}
\bibliography{bib}